\documentclass[10pt,twocolumn,twoside]{IEEEtran}
 \IEEEoverridecommandlockouts                              
\usepackage{amsmath}
\usepackage{arydshln}
\usepackage{mathtools}
\usepackage{amsfonts}
\usepackage{amssymb}
\usepackage{algorithm} 
\usepackage{algpseudocode}
\usepackage{color}
\usepackage{mathrsfs}
\usepackage{stfloats}
\usepackage{balance}
\usepackage[normalem]{ulem}
\usepackage{multirow}

\newtheorem{definition}{Definition}
\newtheorem{lemma}{Lemma}
\newtheorem{remark}{\bf Remark}

\newtheorem{theorem}{Theorem}
\newtheorem{corollary}{Corollary}
\newtheorem{proposition}{Proposition}

\newcommand{\xddots}{%
  \raise 4pt \hbox {.}
  \mkern 10mu
  \raise 1pt \hbox {.}
  \mkern 10mu
  \raise -2pt \hbox {.}
}

\newcommand{\xddotswide}{%
  \raise 5pt \hbox {.}
  \mkern 10mu
  \raise 2pt \hbox {.}
  \mkern 10mu
  \raise -1pt \hbox {.}
}

\title{\LARGE \bf Distributed Synthesis of Local Controllers for Networked Systems with Arbitrary Interconnection Topologies}
\author{Etika Agarwal\textsuperscript{*}, S. Sivaranjani\textsuperscript{*}, Vijay Gupta, and Panos Antsaklis
\thanks{The authors are with the Department of Electrical Engineering, University of Notre Dame,  Notre Dame, IN. \{eagarwal,sseethar,vgupta2,pantsakl\}@nd.edu.
\newline
Funding: E. Agarwal and P. Antsaklis were funded by ARO under grant no. ARL W911NF-17-1-0072. The work of S. Sivaranjani
was supported in part by NSF grants CNS-1544724 and ECCS-1550016, and
that of V. Gupta by NSF grant CNS-1739295, DARPA grant FA8750-20-2-0502 and ARO grant W911NF1910483.\newline
\textsuperscript{*}These authors contributed equally to this work.
}
}

\begin{document}
\maketitle
\thispagestyle{empty}
\pagestyle{empty}

\begin{abstract}
We consider the problem of designing distributed controllers to guarantee dissipativity of a networked system comprised of dynamically coupled subsystems. We require that the control synthesis is carried out locally at the subsystem-level, without explicit knowledge of the dynamics of other subsystems in the network. We solve this problem in two steps. First, we provide distributed subsystem-level dissipativity analysis conditions whose feasibility is sufficient to guarantee dissipativity of the networked system. % an approach to decompose a dissipativity condition on the networked dynamical system into equivalent conditions on the dissipativity of individual subsystems. 
We then use these conditions to synthesize controllers locally at the subsystem-level, using only the knowledge of the dynamics of that subsystem, and limited information about the dissipativity of the subsystems to which it is dynamically coupled. We show that the subsystem-level controllers synthesized in this manner are sufficient to guarantee dissipativity of the networked dynamical system. We also provide an approach to make this synthesis compositional, that is, when a new subsystem is added to an existing network, only the dynamics of the new subsystem, and information about the dissipativity of the subsystems in the existing network to which it is coupled are used to design a controller for the new subsystem, while guaranteeing dissipativity of the networked system including the new subsystem. Finally, we demonstrate the application of this synthesis in enabling plug-and-play operations of generators in a microgrid by extending our results to networked switched systems. 
\end{abstract}
\section{Introduction}
\IEEEPARstart{C}ontrol of large-scale networked dynamical systems, comprising of several dynamically coupled subsystems, has recently gained prominence due to emerging applications in infrastructure networks. For example, in power networks, new architectures where subsystems comprised of small clusters of renewable generators and loads, known as microgrids, are connected to form the large-scale power grid have been proposed \cite{olivares2014trends}. As another example, subsystems comprising of autonomous vehicles that communicate with each other to travel in close formation (platoons), have been proposed to alleviate congestion and enhance safety in transportation networks \cite{varaiya1993smart}\cite{sadraddini2017provably}. 

In such large-scale networks, decentralized and distributed control approaches to guarantee stability and robustness, have been proposed to decrease communication overhead and computational complexity. In these approaches, the subsystem-level controllers only use states from a subset of neighboring subsystems to determine their control actions, however, the process of designing the controllers is centralized, that is, the knowledge of the dynamics of all subsystems is used in the design \cite{antonelli2013interconnected}-\nocite{wang1973stabilization}\nocite{lau1972decentralized}\nocite{aoki1971some}\nocite{bellman1974large}\nocite{davison1990decentralized}\cite{vidyasagar1980decomposition}. This centralized design process has two drawbacks. Firstly, it may be impractical to assume that the control designer has knowledge of the dynamics of all subsystems in the network. For example, in interconnected microgrid networks, where the internal dynamics of the microgrid are continually changing due to renewable energy and load fluctuations, it is not practical to access the dynamics of all the microgrids. Secondly, the topology and interconnection structure of a network may change due to addition or removal of subsystems. For example, in vehicle platooning applications, vehicles may enter or leave the platoon at any time. In such cases, a centralized design process will necessitate redesign of all controllers in the network, which is neither computationally scalable nor desirable for real-time operation.

Therefore, distributed \textit{synthesis}, where controllers are designed locally at the subsystem-level without explicit knowledge of the dynamics of other subsystems, is the only viable option for the realization of such large-scale networks \cite{langbort2010distributed}\cite{farokhi2014decentralized}. Typically, distributed synthesis of controllers has been carried out using three types of approaches. The first class of approaches relies on either exploiting or inducing weak coupling between subsystems in the network to distribute the synthesis problem \cite{sadraddini2017provably}, \cite{bakule1988decentralized}-\nocite{sezer1986nested}\nocite{sethi1998near}\nocite{sivaranjani2017distributed}\cite{sivaranjani2015localization}. The second class of approaches are based on using numerical techniques like methods of multipliers, subgradient algorithms or distributed invariant set computations to decompose the control synthesis problem into more tractable problems \cite{langbort2004distributed}-\nocite{conte2012distributed}\nocite{zeilinger2013plug}\nocite{riverso2014plug}\cite{CUBUKTEPE2018115}. The final class of approaches is hierarchical, involving a centralized computation of subsystem-level conditions to guarantee network-level control objectives such as stability, robustness or dissipativity, and local synthesis of controllers at the subsystem-level to guarantee these objectives \cite{varutti2012dissipativity}-\nocite{ishizaki2019modularity}\nocite{xu2009distributed}\nocite{hudon2012dissipativity}\nocite{tan1990decentralized}\cite{tippett2013dissipativity}. %; however, the dissipativity conditions that the subsystem-level controllers are required to satisfy are computed in a centralized manner. %Alternatively, some dissipativity-based designs employ the fact that negative feedback interconnections of dissipative systems are dissipative, and under mild assumptions, stable; the networked dynamical system is decomposed into negative feedback interconnections of subsystems, and subsystem-level controllers are designed to make each subsystem dissipative .  %\nocite{langbort2006distributed}

In this paper, we consider the problem of synthesizing distributed controllers to guarantee dissipativity  for a networked system comprised of dynamically coupled subsystems. We require that the controllers be designed locally at the subsystem-level without explicit knowledge of the dynamics of other subsystems in the network. The contributions of this paper in addressing this problem are as follows.
\begin{itemize}
    \item \textbf{Distributed analysis:} We first decompose a centralized dissipativity analysis condition on the networked dynamical system into conditions on the dissipativity of individual subsystems. Passivity analysis of a networked system comprised of dynamically coupled subsystems has been studied for star-shaped and cyclical symmetries \cite{wu2011passivity}\cite{ghanbari2016large}, as well as more general interconnection topologies \cite{arcak2016networks}-\nocite{vidyasagar1979new}\nocite{moylan1979tests}\cite{pota1993stability}. However, the passivity verification for the networked system is centralized in these approaches, requiring information about the passivity of all subsystems. In contrast, we propose distributed dissipativity analysis conditions at the subsystem-level, whose feasibility is sufficient to guarantee dissipativity of the networked system. The subsystem-level conditions use only the knowledge of the subsystem dynamics and information about the dissipativity of its neighbors. Further, we do not impose any conditions on the network topology or homogeneity of the subsystem dynamics. 
    \vspace{0.5em}
    \item \textbf{Distributed synthesis:} Using the distributed dissipativity analysis conditions, we then formulate a distributed procedure to synthesize local controllers
     at the subsystem-level to guarantee dissipativity of the networked system. The control synthesis is distributed in the sense that subsystems only use information about their dynamics and the dissipavity of the subsystems to which they are dynamically coupled, to design local subsystem-level controllers.
      \vspace{0.5em}
    \item \textbf{Compositionality:} Finally, we propose an approach to design local controllers for networked systems which may be expanded by adding subsystems at a later stage. When a new subsystem is connected to the networked dynamical system, we formulate a control synthesis procedure that is compositional, that is, the design procedure uses only the knowledge of the dynamics of the newly added subsystem, and the dissipativity of its neighboring subsystems in the existing network, to synthesize local control inputs for the new subsystem, such that the new networked system is dissipative. This procedure does not require redesigning the existing controllers in the network when a new subsystem is added. %The local synthesis of decentralized controllers for dynamically growing interconnections was proposed in \cite{tan1990decentralized}, but the design procedure requires knowledge of the dynamics of all susbystems, which is in contrast to our approach.
     \vspace{0.5em}
\end{itemize}
In addition, we describe how the proposed synthesis approach can be extended to networks of switched systems. Such networks are encountered in several practical applications such as microgrids, where the dynamics and coupling between subsystems change with the availability of renewable generators. Therefore, extending our approach to this setting expands the applicability of our results to a larger class of applications. 
We illustrate the step-by-step implementation of the proposed distributed synthesis through a numerical example, and provide a case study demonstrating the application of this technique in enabling plug-and-play of generators in microgrids.

In \cite{agarwalsequential}, we proposed a preliminary version of this approach to guarantee passivity for a limited class of networked systems with a cascade interconnection topology.  In this paper, we consider arbitrary network topologies, as well as a more general quadratic dissipativity framework, which allows us to capture a variety of properties of interest, such as, $\mathcal{L}_2$ stability, sector-boundedness, conicity, as well as passivity and its variants. We further extend the approach to networks of switched systems, which were not considered in \cite{agarwalsequential}.

This paper is organized as follows. In Section \ref{sec:sys_dyn}, we describe the model of a networked system comprised of dynamically coupled subsystems. We then define dissipativity and formulate the problem of distributed synthesis of local controllers for this system in Section \ref{sec:prob}. In Section \ref{sec:seq}, we present results on the distributed verification of dissipativity, and distributed synthesis of local controllers to guarantee dissipativity of the networked system. In Section \ref{sec:Example}, we present a step-by-step illustration of the proposed  synthesis approach on a numerical example. We then extend these synthesis results to switched systems in Section \ref{sec:switched}, and demonstrate an application to microgrids in Section \ref{sec:case}. The proofs of all the results in this paper are collected in the Appendix. 

\vspace{0.5em}
\textit{Notation:} We denote the sets of real numbers, positive real numbers including zero, and $n$-dimensional real vectors by $\mathbb{R}$, $\mathbb{R}^{+}$ and $\mathbb{R}^{n}$ respectively. Define $\mathbb{N}_N=\{1,\ldots,N\}$, where $N$ is a natural number excluding zero. Given a block matrix $A=[A_{i,j}]_{i\in\mathbb{N}_n,j\in\mathbb{N}_m}$, $A_{i,j}$ represents the $(i,j)$-th block, and $A'\in \mathbb{R}^{n \times m}$ represents its transpose. Given matrices $A_{1},\ldots, A_{i}$, $\mbox{diag}(A_{1},\ldots,A_{i})$ represents a block-diagonal matrix with $A_{1},\ldots, A_{i}$ as its diagonal entries. A symmetric positive definite matrix $P \in \mathbb{R}^{n \times n}$ is represented as $P>0$ (and as $P\geq 0$, if it is positive semi-definite). The standard identity matrix is denoted by $\mathbf{I}$, with dimensions clear from the context. Given sets $A$ and $B$, $A\backslash B$ represents the set of all elements of $A$ that are not in $B$. 

\section{System Dynamics}\label{sec:sys_dyn}

Consider a networked dynamical system $\mathbf{T_N}$ comprised of $N$ subsystems, as shown in Fig. 1, where the dynamics of the $i$-th subsystem $\Sigma_i$, $i \in \mathbb{N}_N$ is described by
\begin{equation}\label{linear system}
\begin{aligned}
\dot{x}_i(t) & = A_i x_i(t) + B_i^{(1)}v_i(t) + B_i^{(2)} w_i(t) + B_i^{(3)} u_i(t), \\
y_i(t) & = C_ix_i(t), \\
v_i(t) & = \sum \limits_{j \in \mathbb{N}_N} H_{i,j} x_{j}(t),
\end{aligned}
\end{equation}

\noindent where \(x_i(t) \in \mathbb{R}^{n_i} \), \(y_i(t) \in \mathbb{R}^{m_i}\),   \(v_i(t)\in \mathbb{R}^{z_i}\), \(w_i(t)\in\mathbb{R}^{l_i}\) and \(u_i(t)\in \mathbb{R}^{p_i}\) are the state, output, coupling input, exogenous disturbance, and control input respectively. 

A subsystem $\Sigma_j$, $j \in \mathbb{N}_N\backslash\{i\}$ is said to be dynamically coupled with the subsystem $\Sigma_i$ if $H_{i,j}\neq 0$. Define the neighbor set for $\Sigma_i$ to be $$\mathcal{E}_i=\{j: H_{i,j}\neq 0, j \in \mathbb{N}_N\backslash\{i\} \}.$$

The dynamics of the networked system $\mathbf{T_N}$ is written as
   \begin{equation}\label{interconnected_linear}
\begin{aligned}
	\dot{x}(t) & = Ax(t) + B^{(1)}v(t) + B^{(2)}w(t) + B^{(3)}u(t)\\
	y(t) & = Cx(t)\\
	v(t) & = Hx(t)
	\end{aligned}
	\end{equation}
\noindent	where 
	\begin{align*}
	A &=\mbox{diag}(A_1, A_2, \dots, A_N)\\
	B^{(j)} &=\mbox{diag}(B^{(j)}_1, B^{(j)}_2,\dots,B^{(j)}_N), \quad j\in \mathbb{N}_3,\\
	C &=\mbox{diag}(C_1, C_2, \dots, C_N) \\
	H  &=  \begin{bmatrix} \nonumber 
	H_{i,j}
	\end{bmatrix}_{i,j \in \mathbb{N}_N},
	\end{align*}
	and \(x(t) \in \mathbb{R}^{n} \), \(y(t) \in \mathbb{R}^{m}\),   \(v(t)\in \mathbb{R}^{z}\), \(u(t)\in \mathbb{R}^{p}\), and \(w(t)\in\mathbb{R}^{l}\) are the augmented system state, output, coupling input, control input, and disturbance formed by stacking $x_{i}(t)$, $y_{i}(t)$, $v_{i}(t)$, $u_{i}(t)$, and $w_{i}(t)$ respectively of all $N$ subsystems.  The matrix $H$ represents the dynamical coupling in the system, and is referred to as the coupling matrix. 

We denote the interconnected system $\mathbf{T_{N+1}}$ obtained by connecting a new subsystem $\Sigma_{N+1}$ to $\mathbf{T_N}$ by $\mathbf{T_{N+1}}:=\mathbf{T_N}\vert\Sigma_{N+1}$. 
\begin{figure}[t]
	\centering
	%%\vspace{-1.4em}
	\includegraphics[scale=0.63]{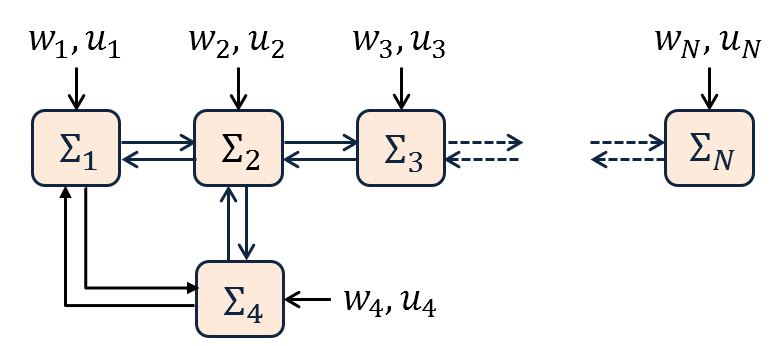}
	\caption{Example of a networked dynamical system $\mathbf{T_N}$.}
	\label{fig:cascade}
	% %%\vspace{-0.1em}
\end{figure}

\section{Problem Description}\label{sec:prob}
In this section, we formulate the problem of synthesizing local controllers at the subsystem-level in a distributed manner to enforce a quadratic dissipativity property on the interconnected system. 
 \vspace{0.5em}
\begin{definition}\cite{nijmeijer1993passive}\label{def:diss}
 A dynamical system \eqref{interconnected_linear} is said to be $QSR$-dissipative from $w$ to $y$, if there exists a positive definite function $V(x):\mathbb{R}^n\longrightarrow\mathbb{R}^+$, called the storage function, such that, for all $t>t_0\geq0$, $x(t_0)\in\mathbb{R}^n$, and $w(t)\in\mathbb{R}^l$,
 
 \begin{align}\label{eq:qsr}
     \int_{t_0}^{t} \begin{bmatrix}y(\tau) \\ w(\tau)\end{bmatrix}'\begin{bmatrix}Q & S \\ S' & R\end{bmatrix} \begin{bmatrix}y(\tau) \\ w(\tau)\end{bmatrix} d\tau\geq V(x(t))-V(x(t_0))
 \end{align}
 \vspace{0.5em}
 
 \noindent holds, where $x(t)$ is the state at time $t$ resulting from the initial condition $x(t_0)$, and $Q$, $S$ and $R$ are matrices of appropriate dimension. \end{definition}
 \vspace{0.5em}
 We consider the objective of enforcing $QSR$-dissipativity on the networked system, as it can be used to capture a wide variety of dynamical properties of interest through appropriate choices of the $Q$, $S$ and $R$ matrices as follows. 
 \vspace{0.5em}
\begin{remark}\label{rem:cases} \cite{forbes2011extensions}
System \eqref{interconnected_linear} satisfying Definition \ref{def:diss} is,
\begin{enumerate}
    \item \textit{passive}, if $Q = 0$, $S = \frac{1}{2}I$ and $R =0$,
    \item \textit{strictly passive}, if $Q=-\rho I$, $S= \frac{1}{2}I$ and $R =-\nu I$, where $\rho,\nu \in \mathbb{R}^+\backslash\{0\}$,
    \item \textit{$\mathcal{L}_2$ stable}, if $Q=-\frac{1}{\gamma} I$, $S= 0$ and $R =\gamma I$ where $\gamma\in \mathbb{R}^+$ is an $\mathcal{L}_2$ gain of the system, 
    \item \textit{conic}, if $Q=-I$, $S = c I$ and $R = (r^2-c^2)I$, where $c\in\mathbb{R}$ and $r \in \mathbb{R}^+\backslash\{0\}$, and,
    \item \textit{sector-bounded}, if $Q=-I$, $S = \frac{a+b}I$ and $R = -ab I$, where $a,b\in \mathbb{R}$.
\end{enumerate}
\end{remark}
 \vspace{0.5em}
Note that the $QSR$-dissipativity condition \eqref{eq:qsr} is a special case of an integral quadratic constraint (IQC) on $(w,y)$, where the multiplier is an identity
matrix \cite{megretski1997system}. Under mild assumptions, the $QSR$-dissipativity property can also be used to guarantee Lyapunov stability \cite{kottenstette2014relationships}. $QSR$-dissipativity of the dynamical system \eqref{interconnected_linear} can be analyzed as follows.
 \vspace{0.5em}
\begin{proposition}[Centralized dissipativity analysis]\label{prop:central_analysis}\cite{kottenstette2014relationships}
A dynamical system \eqref{interconnected_linear} is $QSR$-dissipative with $V(x)=x'Px$ if there exists a positive definite matrix $P>0$ and matrices $Q$, $S$ and $R$ of appropriate dimensions such that

\begin{equation}\label{eq:central_analysis}
    \Gamma = \begin{bmatrix}
    -\hat{A}'P-P\hat{A}'+C'QC & -PB^{(2)}+C'S \\
    -B^{(2)'}P+S'C & R
    \end{bmatrix} \geq 0
\end{equation}

\noindent holds, where $\hat{A} = A + B^{(1)}H$.
\end{proposition}
 \vspace{0.5em}

The analysis condition in Proposition \ref{prop:central_analysis} is centralized in the sense that a solution to \eqref{eq:central_analysis} requires knowledge of the dynamics and coupling of all subsystems in the networked dynamical system $\mathbf{T_N}$. However, for large-scale networks where new subsystems may be added or removed, it is desirable to develop an analysis and control synthesis that can be carried out locally at the subsystem-level with limited knowledge of the dynamics and coupling with neighboring subsystems. In this context, the aim of this paper is to address the following problems:
\begin{enumerate}
	 \vspace{0.5em}
    \item \textbf{Distributed analysis:} Decompose the analysis condition in Proposition \ref{prop:central_analysis} into conditions on the dissipativity of subsystems $\Sigma_i, i \in \mathbb{N}_N$.
     \vspace{0.5em}
    \item \textbf{Distributed synthesis:} Formulate a procedure to design local control inputs 
      \vspace{0.5em}
     \begin{align}
    	u_i(t)&=\sum \limits_{j \in \mathcal{E}_i \cup \{i\}}u_{i,j}(t), \, &i \in \mathbb{N}_N,\\
    	u_{i,j}(t)&= K_{i,j}x_j(t), \,  & j \in \mathcal{E}_i \cup \{i\},
    \end{align}
      \vspace{0.2em}
      
  \noindent such that $\mathbf{T_N}$ is $QSR$-dissipative, where the synthesis of the controller matrices $K_{i,j}$ only uses the dynamics \eqref{linear system}, and information about the dissipativity of its neighboring subsystems $\Sigma_j, j \in \mathcal{E}_i$.
     \vspace{0.5em}
    \item \textbf{Compositionality:} When a new subsystem $\Sigma_{N+1}$ is connected to the networked dynamical system $\mathbf{T_N}$, obtain a control synthesis procedure which is compositional, that is, the design procedure uses only the knowledge of the dynamics of $\Sigma_{N+1}$, and the dissipativity of its neighboring subsystems $\Sigma_j, j \in \mathcal{E}_{N+1}$, to synthesize local control inputs $u_{j,N+1}(t), j\in \mathcal{E}_{N+1},$ and 
    
    $$u_{N+1}(t)=\sum \limits_{j \in \mathcal{E}_{N+1} \cup \{N+1\}} u_{N+1,j}(t),$$ 
    
    \noindent such that $\mathbf{T_{N+1}}:=\mathbf{T_N}\vert\Sigma_{N+1}$ is $QSR$-dissipative. 
     \vspace{0.5em}
\end{enumerate}

%%%%%%%%%%%%%%%%%%%%%%%%%%%%%%%%%%%%%%%%%%%%%%%%%%%%%%%%
\section{Distributed Synthesis of Local Controllers}\label{sec:seq}
In this section, we present a distributed approach to synthesize local (subsystem-level) controllers that guarantee dissipativity of the networked dynamical system $\mathbf{T_N}$.  In this approach, every subsystem synthesizes a local controller using only the knowledge of its own dynamics, and information about the dissipativity of the subsystems to which it is dynamically coupled. The proofs of all the results in this section are collected in the Appendix.

\begin{figure*}[b]
	\setcounter{equation}{9}
	%%\vspace{0.5em}
		\hrulefill
	\begin{subequations}\label{eq:messenger_matrix}
		\begin{align}
		\mathcal{M}_i &= \begin{cases} \mu^s_1, & i =1  \\ \mu^s_i - \mu^c_i, & i \in \mathbb{N}_N \backslash \{1\}
		\end{cases} \\
		% Q & = diag (Q_1, Q_2, \ldots, Q_N) \\
		\hat{H}_{i,j} &= P_iB_i^{(1)}H_{i,j}, \hspace{25.7em} i\in\mathbb{N}_N,\, j \in \mathcal{E}_i \cup\{i\} \\
		\mu^s_i & = \begin{bmatrix} -(A'_i P_i+P_i A_i)-(\hat{H}_{i,i}+\hat{H}'_{i,i}) + C_i'Q_i C_i & -P_i B_i^{(2)}+C'_i S_i \\
		-B_i^{(2)'}P_i+S'_i C_i & R_i\end{bmatrix}, \hspace{3.2em} i \in \mathbb{N}_N
		\\\mu^c_i &= \sum \limits_{j\in \mathcal{E}_i \cap \mathbb{N}_{i-1}}\begin{bmatrix}\hat{H}'_{j,i}+\hat{H}_{i,j} & 0\\ 0 & 0 \end{bmatrix} \left(\mathcal{M}_{j}\right)^{-1}\begin{bmatrix}\hat{H}'_{j,i}+\hat{H}_{i,j} & 0\\ 0 & 0 \end{bmatrix}', \hspace{7.5em} i \in \mathbb{N}_N\backslash \{1\}.
		\end{align}
	\end{subequations}
	%%\vspace{-1em}
	% \setcounter{equation}{\value{MYtempeqncnt}}
\end{figure*}
\setcounter{equation}{6}

\subsection{Distributed Analysis}
We begin by distributing the dissipativity analysis condition in Proposition \ref{prop:central_analysis}. We derive a property of positive definite matrices that will be useful in this context. 
  \vspace{0.5em}
\begin{lemma}\label{lemma:Cholesky_blk}
A symmetric block matrix

\begin{equation}\label{eq:block_matrix}
    W = \begin{bmatrix} 
   W_{1,1} & W_{1,2} & \dots & W_{1,N} \\
   W_{2,1} & W_{2,2} & \dots & W_{2,N} \\
   \vdots & \vdots & & \vdots \\
   W_{N,1} & W_{N,2} & \dots & W_{N,N} \\
    \end{bmatrix},
    \vspace{0.3em}
\end{equation}

\noindent where $W_{i,j}$, $i,j \in \mathbb{N}_N$ are block matrices of appropriate dimension, is positive definite if and only if

    \begin{equation}\label{seq_Cholesky_lemma}
    \begin{aligned}
    {M}_i & > 0,\quad  \forall i \in \mathbb{N}_N,  \\
        {M}_i & =\begin{cases} W_{i,i}, & i =1  \\  W_{i,i} - \sum\limits_{k=1}^{i-1} W_{i,k}{M}_{k}^{-1} W_{k,i}, & i \in \mathbb{N}\backslash\{1\}.
        \end{cases}
        \end{aligned}
    \end{equation}
    
\end{lemma}
\vspace{0.5em}

The condition \eqref{seq_Cholesky_lemma} allows for the verification of the positive definiteness of a matrix to be carried out row-wise. Now, observe that the dissipativity of a networked dynamical system $\mathbf{T_N}$ can be analyzed by ascertaining the positive definiteness of matrix $\Gamma$ in \eqref{eq:central_analysis}. We can then use Lemma \ref{lemma:Cholesky_blk} to decompose \eqref{eq:central_analysis} into conditions that can be verified at the subsystem-level. We have the following result.
 \vspace{0.5em}
\begin{theorem}[Distributed dissipativity analysis] \label{thm:seq_analysis}
The networked system $\mathbf{T_N}$ \eqref{interconnected_linear} is $QSR$-dissipative from $w$ to $y$ with

\begin{align*}
Q &= \mbox{diag}(Q_1, Q_2, \dots, Q_N), \quad Q_i~\in~\mathbb{R}^{m_i \times m_i}\\
S &= \mbox{diag}(S_1, S_2, \dots, S_N), \quad S_i \in \mathbb{R}^{m_i \times l_i}\\
R &= \mbox{diag}(R_1, R_2, \dots, R_N), \quad R_i\in \mathbb{R}^{l_i \times l_i},\\
\end{align*}

\noindent $i \in \mathbb{N}_N$, if there exist matrices $P_i$,  termed \textit{energy matrices}, such that

%and matrices $Q_i$, $S_i$  and $R_i$, $i\in\mathbb{N}_N$, such that 
\begin{equation}\label{seq_Cholesky}
\begin{aligned}
    \mathbb{P}_1:   \mbox{Find } \quad & P_i \\
%    , Q_i, S_i, R_i\\
     \mbox{s.t.} \quad & P_i>0, \ \mathcal{M}_i  > 0, \\
        & P_i \in \mathbb{R}^{n_i \times n_i}
%     Q_i \in \mathbb{R}^{m_i \times m_i}, \\
%    & S_i \in \mathbb{R}^{m_i \times l_i}, \,
%     R_i\in \mathbb{R}^{l_i \times l_i}
\end{aligned}
\end{equation}

\noindent is feasible $\forall i\! \in\!\mathbb{N}_N$, where $\mathcal{M}_i$ is computed in \eqref{eq:messenger_matrix}.
\end{theorem}

\begin{remark}[Messenger matrix]\label{rem:messengermatrix}
Theorem \ref{thm:seq_analysis} provides distributed subsystem-level conditions for the verification of network-level dissipativity. The centralized analysis condition \eqref{eq:central_analysis} is decomposed into local conditions \eqref{seq_Cholesky}, where each subsystem computes and stores an information matrix called as the \textbf{\textit{messenger matrix}} $\mathcal{M}_i$, $i \in \mathbb{N}_N$. The messenger matrix for each subsystem $\Sigma_i$, $i \in \mathbb{N}_N$, is the difference between two terms, (i) $\mu_i^s$, which can be interpreted as the dissipativity of the subsystem, and (ii) $\mu_i^c$, which can be interpreted as the energy flow from its neighbors. The term $\mu_i^c$ contains information about the dynamical coupling between the subsystem $\Sigma_i$ and its neighbors, as well as aggregated information about the dissipativity of its neighboring subsystems through the messenger matrices $\mathcal{M}_j$, and and energy matrices $P_j$, $j \in \mathcal{E}_i \cap \mathbb{N}_{i-1}$, which are communicated by neighbors $\Sigma_{j}$, $j \in \mathcal{E}_i \cap \mathbb{N}_{i-1}$. The positive definiteness of all messenger matrices (which can be verified at the subsystem-level) is sufficient to guarantee the dissipativity of the networked system $\mathbf{T_N}$. 
\end{remark}
% \vspace{0.5em}

The analysis in Theorem \ref{thm:seq_analysis} can be implemented as described in Algorithm \ref{algo:seq_analysis}.
\begin{algorithm}[H]
	\caption{Distributed Analysis for $\mathbf{T_N}$}
	\label{algo:seq_analysis}
	\begin{algorithmic}[1]
		\State Initialize $i = 1$. %, $\mathcal{M}^{cl}_0 = 0$. %$\mathcal{M}_0 = 0$,
		\While{$i\leq N$, at subsystem $\Sigma_i$,}\label{loop}
		\If{$i\neq1$}
		\State Receive $\mathcal{M}_{j}$ and $P_j$ from $\Sigma_{j}$, $j\in\mathcal{E}_i \cap \mathbb{N}_{i-1}$.
		% \State Set $\mathcal{M}_{i-1} = \mathcal{M}^{cl}_{i-1}$.
		\EndIf
		\If{$\mathbb{P}_1$ is feasible}
		\State Compute $\mathcal{M}_i>0$ and $P_i>0$ from \eqref{eq:messenger_matrix}. % and \eqref{messenger_matrix} and set $\mathcal{M}^{cl}_i = \mathcal{M}_i$.
%		\State Set $K_{i,i}=K_{i,j}=K_{j,i}=0$, $\forall j \in \mathcal{E}_{i}$.
		\Else
		\State Return ``\textit{infeasible}''.
		\State Go to Step \ref{end}.
%		\State{\textit{Control design:}} Solve $\mathbb{P}_2$ to compute $K_{i,i}$, $K_{i,j}$, $K_{j,i}$, $j\in\mathcal{E}_i$ and $\mathcal{M}_i>0$ from  \eqref{eq:messenger_matrix_control}.
		\EndIf
%		\State Send $K_{j,i}$ to $\Sigma_{j}$, $j\in\mathcal{E}_i$.
		\State Set $i\mapsto i+1$.
		\EndWhile \label{end}
	\end{algorithmic}
\end{algorithm}
\begin{figure*}[b]
    \setcounter{equation}{13}
    \hrulefill
    \begin{subequations}\label{eq:messenger_matrix_control}
    \begin{align}
    \mathcal{M}_i &= \begin{cases} \mu^s_1, & i =1  \\ \mu^s_i - \mu^c_i, & i \in \mathbb{N}_N \backslash \{1\}
        \end{cases} \\
    % Q & = diag (Q_1, Q_2, \ldots, Q_N) \\
    \hat{H}_{i,j} &= P_i(B_i^{(1)}H_{i,j}+B_i^{(3)}K_{i,j}), \hspace{20.2em} i\in\mathbb{N}_N,\, j \in \mathcal{E}_i \cup\{i\} \\
    \mu^s_i & = \begin{bmatrix} -(A'_i P_i+P_i A_i)-(\hat{H}_{i,i}+\hat{H}'_{i,i}) + C_i'Q_i C_i & -P_i B_i^{(2)}+C'_i S_i \\
    -B_i^{(2)'}P_i+S'_i C_i & R_i\end{bmatrix}, \hspace{3.4em} i \in \mathbb{N}_N
    \\\mu^c_i &= \sum \limits_{j\in \mathcal{E}_i \cap \mathbb{N}_{i-1}}\begin{bmatrix}\hat{H}'_{j,i}+\hat{H}_{i,j} & 0\\ 0 & 0 \end{bmatrix} \left(\mathcal{M}_{j}\right)^{-1}\begin{bmatrix}\hat{H}'_{j,i}+\hat{H}_{i,j} & 0\\ 0 & 0 \end{bmatrix}', \hspace{7.7em} i \in \mathbb{N}_N \backslash \{1\}.
    \end{align}
\end{subequations}
\end{figure*}
\setcounter{equation}{10}
\begin{remark}\label{rem:messenger_compuation} We make the following remarks about the computation of the messenger matrix used in Theorem \ref{thm:seq_analysis}.
\begin{itemize}
    \item[(i)] The computation of messenger matrix $\mathcal{M}_i$ in \eqref{eq:messenger_matrix} requires $\mathcal{M}_j^{-1}$, $j \in \mathcal{E}_i\cap \mathbb{N}_{i-1}$. However, in cases where $R_j=0$, $\mathcal{M}_j$ takes the form
	$\mathcal{M}_j= \left [ \begin{array}{c;{2pt/2pt}c}
	a_{1,j} & a_{2,j} \\ \hdashline[2pt/2pt]
	a'_{2,j} & 0
	\end{array}\right ].$
	Then, $\mathcal{M}_i$ can be computed by replacing the expression for $\mathcal{M}_j^{-1}$ in \eqref{eq:messenger_matrix} by
	$\mathcal{M}_j^{-1}= \left [ \begin{array}{c;{2pt/2pt}c}
	a_{1,j}^{-1} & 0 \\ \hdashline[2pt/2pt]
	0 & 0
	\end{array}\right ],$
	 and relaxing the condition $\mathcal{M}_j>0$ to $\mathcal{M}_j\geq 0$ in $\mathbb{P}_1$. Note that this holds for all the results to follow, which will involve computation of messenger matrix.
 \item[(ii)] If the dynamics of subsystem $\Sigma_i$ includes a feedthrough term $D_i \in \mathbb{R}^{m_i\times l_i}$ such that $y_i(t)=C_ix_i(t)+D_iw_i(t)$, then the computation of the messenger matrix $\mathcal{M}_i$ in \eqref{eq:messenger_matrix} can be modified as follows. The term $R_i$ in \eqref{eq:messenger_matrix}-(c) is modified to $R_i \mapsto \hat{R}_i$, where $\hat{R}_i= -D_i'Q_iD_i-(D_i'S_i+S_i'D_i)$. 
\end{itemize}
\end{remark}
\vspace{1em}
\begin{remark}\label{rem:sequential}
In the $i$-th iteration of Algorithm \ref{algo:seq_analysis} (Steps 3-11), the dissipativity of the network $\mathbf{T_i}$ formed by the interconnection of subsystems $\Sigma_1,\Sigma_2,\cdots,\Sigma_i$ is verified. Therefore, the messenger matrices $\mathcal{M}_i$, $i \in \mathbb{N}_N$ will vary with the choice of numbering assigned to subsystems in the network, and the distributed analysis conditions in Theorem \ref{thm:seq_analysis} are only sufficient to guarantee dissipativity of the networked system $\mathbf{T_N}$.
\end{remark}

\begin{remark}[Robustness to modeling uncertainties]
In practice, the system matrices $A_i$, $B_i$, and $C_i$ are known upto a certain level of modeling uncertainty. Since the proposed distributed analysis conditions only require the messenger matrix $\mathcal{M}_i$ to be positive definite, it is possible to include a non-zero lower bound on $\mathcal{M}_i$ which will account for uncertainties in the system model. For example, assume $A_i$ has an additive uncertainty $\Delta A_i$, that is, $A_i \mapsto A_i+\Delta A_i$. Then, the robustness margin for $\mathcal{M}_i$ can be derived as follows. The term $\mu^s_i$ in \eqref{eq:messenger_matrix}-(c) is updated to $\hat{\mu}^s_i$, where $\hat{\mu}^s_i = \mu^s_i + \left [ \begin{array}{c;{2pt/2pt}c}
	-(\Delta A'_i P_i+P_i \Delta A_i) & 0 \\ \hdashline[2pt/2pt]
	0 & 0
	\end{array}\right ]$. The messenger matrix in Theorem \ref{thm:seq_analysis} is now updated to $$\hat{\mathcal{M}}_i = \mathcal{M}_i + \Bigg[ \begin{array}{c;{2pt/2pt}c}
	-(\Delta A'_i P_i+P_i \Delta A_i) & 0 \\ \hdashline[2pt/2pt]
	0 & 0
	\end{array}\Bigg], $$
	and the verification condition $\mathcal{M}_i>0$ to $\hat{\mathcal{M}}_i>0$, or 
	\begin{align} \label{eq:M_robust}
	\mathcal{M}_i > \Bigg[ \begin{array}{c;{2pt/2pt}c}
	(\Delta A'_i P_i+P_i \Delta A_i) & 0 \\ \hdashline[2pt/2pt]
	0 & 0
	\end{array}\Bigg].
	\end{align}
	If $\Delta A_i$ is norm bounded, that is,  $\|\Delta A_i\| < \epsilon_i$ then updating the verification condition $\mathcal{M}_i>0$ in Theorem \ref{thm:seq_analysis} to $\mathcal{M}_i > \mathcal{M}_i^{bound} = \Bigg[ \begin{array}{c;{2pt/2pt}c}
	2\epsilon_i P_i & 0 \\ \hdashline[2pt/2pt]
	0 & 0
	\end{array}\Bigg]$ will ensure \eqref{eq:M_robust}, and hence the robustness of the proposed distributed analysis conditions to the additive uncertainty. The derivation of $\mathcal{M}_i^{bound}$ for other classes of uncertainties like multiplicative or parametric uncertainties will require extending the proposed framework to a general integral quadratic constraint (IQC)  setup \cite{megretski1997system} , and is a subject of future work.
\end{remark}

\subsection{Distributed Synthesis}
Theorem \ref{thm:seq_analysis} provides sufficient conditions at the subsystem-level to guarantee dissipativity of the networked dynamical system $\mathbf{T_N}$. If the dissipativity conditions in Theorem \ref{thm:seq_analysis} are not met, we would like to synthesize local controllers at the subsystem-level to guarantee dissipativity of the networked system. Further, we require that the control synthesis be carried out at the subsystem-level, using only the dynamics of the subsystem and the messenger matrices communicated from its neighbors. We have the following result on distributed synthesis.
\vspace{0.5em}
\begin{theorem}\label{thm:distributed_synthesis}
 The local control inputs %\setcounter{equation}{10}
 
 \begin{equation}
      \begin{aligned}
     &u_i(t)=\sum \limits_{j \in \mathcal{E}_i \cup \{i\}}u_{i,j}(t), \; i \in \mathbb{N}_N,\\
     &u_{i,j}(t)= K_{i,j}x_j(t), \; j \in \mathcal{E}_i \cup \{i\},
 \end{aligned}
 \end{equation}
 
\noindent designed by solving
\vspace{0.6em}
\begin{equation}\label{close_loop_seq_Cholesky}
\begin{aligned}
    \mathbb{P}_2:   \mbox{Find } \quad & P_i, K_{i,i}, K_{i,j}, K_{j,i}, \quad j \in \mathcal{E}_i \\
    %Q_i, S_i, R_i,
     \mbox{s.t.} \quad & P_i>0, \\
     & \mathcal{M}_i  > 0, \\
    % & \left [ \begin{array}{c;{2pt/2pt}c}
    % \mu_i^s & \begin{array}{c;{2pt/2pt}c}
    % \end{array} \\ \hdashline[2pt/2pt]
    % -B^{(2)'}P+S'C & R
    % \end{array}\right ]
    % \begin{bmatrix}
    % \mu_i^s & 
    % \end{bmatrix}
      & P_i \in \mathbb{R}^{n_i \times n_i}, \\
   & K_{i,i}  \in \mathbb{R}^{p_i \times n_i}, \\
        & K_{i,j} \in \mathbb{R}^{p_i \times n_j}, \, K_{j,i} \in \mathbb{R}^{p_j \times n_i},
        %Q_i \in \mathbb{R}^{m_i \times m_i}, \\
       % & S_i \in \mathbb{R}^{m_i \times l_i}, \,
        %R_i\in \mathbb{R}^{l_i \times l_i}, \\
\end{aligned}
\end{equation}
\vspace{0.6em}

\noindent for all $i\in\mathbb{N}_N$, where $\mathcal{M}_i$ is the closed-loop messenger matrix of $\Sigma_i$ computed from \eqref{eq:messenger_matrix_control}, render $\mathbf{T_N}$ $QSR$-dissipative with
\vspace{0.2em}
\begin{align*}
Q &= \mbox{diag}(Q_1, Q_2, \dots, Q_N), \quad Q_i~\in~\mathbb{R}^{m_i \times m_i}\\
S &= \mbox{diag}(S_1, S_2, \dots, S_N), \quad S_i \in \mathbb{R}^{m_i \times l_i}\\
R &= \mbox{diag}(R_1, R_2, \dots, R_N), \quad R_i\in \mathbb{R}^{l_i \times l_i}, \quad i \in \mathbb{N}_N.\\
\end{align*}
\end{theorem}

The synthesis in Theorem \ref{thm:distributed_synthesis} can be carried out as described in Algorithm \ref{algo:seq_synthesis}.

\begin{algorithm}[H]
\vspace{0.5em}
	\caption{Distributed Synthesis for $\mathbf{T_N}$}
	\label{algo:seq_synthesis}
	\begin{algorithmic}[1]
	    \State Initialize $i = 1$. %, $\mathcal{M}^{cl}_0 = 0$. %$\mathcal{M}_0 = 0$,
	    \While{$i\leq N$, at subsystem $\Sigma_i$,} \label{loop}
	    \If{$i\neq1$}
	    \State Receive $\mathcal{M}_{j}$ and $P_j$ from $\Sigma_{j}$, $j\in\mathcal{E}_i \cap \mathbb{N}_{i-1}$.
	    % \State Set $\mathcal{M}_{i-1} = \mathcal{M}^{cl}_{i-1}$.
	    \EndIf
	   % \State Set $\mathcal{M}_{i-1} = \mathcal{M}^{cl}_{i-1}$.
	    \If{$\mathbb{P}_1$ is feasible}
	    \State Compute $\mathcal{M}_i>0$ from \eqref{eq:messenger_matrix}. % and \eqref{messenger_matrix} and set $\mathcal{M}^{cl}_i = \mathcal{M}_i$.
	    \State Set $K_{i,i}=K_{i,j}=K_{j,i}=0$, $\forall j \in \mathcal{E}_{i}$.
	    \Else
	    \State{\textit{Control design:}} Solve $\mathbb{P}_2$ to compute $K_{i,i}$, $K_{i,j}$, $K_{j,i}$, $j\in\mathcal{E}_i\cap \mathbb{N}_{i-1}$ and $\mathcal{M}_i>0$, $P_i>0$ from  \eqref{eq:messenger_matrix_control}.
	    \EndIf
	    \State Send $K_{j,i}$ to $\Sigma_{j}$, $j\in\mathcal{E}_i\cap \mathbb{N}_{i-1}$.
		\State Set $i\mapsto i+1$.
		\EndWhile
	\end{algorithmic}
\end{algorithm}

At the $i$-th iteration of Algorithm \ref{algo:seq_synthesis} (Steps 3-12), two types of controller matrices are designed at subsystem $\Sigma_i$ to guarantee the dissipativity of the subnetwork $\mathbf{T_i}$ formed by the interconnection of subsystems $\Sigma_1,\Sigma_2,\cdots,\Sigma_i$ - (i) self controller matrix $K_{i,i}$, and (ii) coupling controller matrices $K_{i,j}$ and $K_{j,i}$, $j\in\mathcal{E}_i\cap \mathbb{N}_{i-1}$, corresponding to the bidirectional interconnections with its neighbors in $\mathbf{T_i}$. Note that the existing controllers in the subnetwork $\mathbf{T_{i-1}}$ are not redesigned at the $i$-th iteration, since the control synthesis at $\Sigma_i$ is carried out to ensure the dissipativity of $\mathbf{T_i}=\mathbf{T_{i-1}}\vert \Sigma_i$. 
\vspace{0.5em}
\begin{remark}
	The energy flow between subsystems, $\mu_i^c, i \in \mathbb{N}_N$  in \eqref{eq:messenger_matrix_control}, need not be small; therefore, the controller does not require or enforce weak coupling between subsystems.
\end{remark}
\vspace{0.5em}

The control synthesis equations in $\mathbb{P}_2$ are bilinear; however, they can readily be expressed as linear matrix inequalities using a Schur's complement method \cite[Section 4.6]{vanantwerp2000tutorial}. We also note that akin to any distributed approach, our synthesis yields more conservative controllers than those obtained by a centralized synthesis.

\begin{remark}[Feasibility of distributed synthesis and local performance]\label{rem:feasibility}
Theorem \ref{thm:distributed_synthesis} relies on the distributed verification of dissipativity presented in Theorem \ref{thm:seq_analysis} to locally design linear subsystem-level controllers that guarantee the dissipativity of the networked system. While the underlying verification problem relies on the sufficient-only conditions in Theorem \ref{thm:seq_analysis} that may not always be feasible, the only consideration for the feasibility of the synthesis problem in Theorem \ref{thm:distributed_synthesis} is the existence of a controller that ensures the positive-definiteness of the messenger matrix in \eqref{eq:messenger_matrix_control}. If all the subsystems in the network are minimum phase and $R_i\geq 0$ (or $\hat{R}_i\geq0$ for systems with feedthrough terms, where $\hat{R}_i$ is defined in Remark \ref{rem:messenger_compuation}), then the linear feedback control design problem in Algorithm \ref{algo:seq_synthesis} is \textit{always} feasible, irrespective of the sequence (numbering) of subsystems used to solve the synthesis problem.  However, since \eqref{eq:messenger_matrix_control} is an LMI that can have multiple solutions, a secondary control objective may be used to choose a controller for a specific application. In this context, it is also possible to incorporate subsystem-level performance objectives in the distributed control synthesis problem while providing global dissipativity guarantees, since the addition of local optimization objectives at the subsystem-level does not affect the feasibility of the synthesis. However, the addition of control constraints may affect the feasibility of the synthesis problem by restricting the space of controllers for which the positive-definiteness of the messenger matrix can be guaranteed.
\end{remark}

\begin{figure}[t]
	\centering
	\vspace{0.5em}
	\includegraphics[scale=0.54,trim=0.1cm 0.6cm 0cm 0cm]{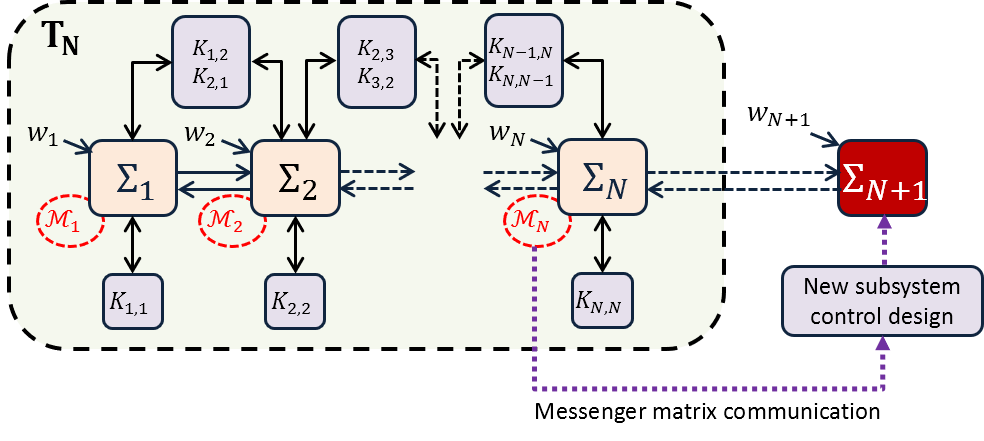}
	\caption{Schematic of compositional control design for $\mathbf{T_{N+1}}=\mathbf{T_N}\vert\Sigma_{N+1}$ when a new subsystem $\Sigma_{N+1}$ is connected to $\mathbf{T_N}$.}
	\label{fig:compositional_control}
	\vspace{-0.5em}
\end{figure}
\subsection{Compositional Analysis and Control Synthesis}\label{sec:compositional}
In large-scale networks, new subsystems may be connected to the existing network at a later time. In such scenarios, it is desirable to guarantee dissipativity of the updated network \textit{compositionally}, that is, without redesigning pre-existing controllers. In this subsection, we extend the synthesis in Theorem~\ref{thm:distributed_synthesis} to design local controllers for the newly added subsystem, using only limited information from its neighbors to guarantee dissipativity of the new networked system.

\begin{figure*}[!t]
%	\small{
		\setcounter{equation}{16}
		%%\vspace{0.5em}
		\begin{subequations}\label{eq:messenger_matrix_comp_control}
			\begin{align}
			\mathcal{M}_{N+1} &= \mu^s_{N+1} - \mu^c_{N+1} \\
			% Q & = diag (Q_1, Q_2, \ldots, Q_N) \\
			\hat{H}_{{N+1},j} &= P_{N+1}(B_{N+1}^{(1)}H_{{N+1},j} +B_i^{(3)}K_{N+1,j}), \hspace{1em} j \in \mathcal{E}_{N+1} \cup\{N+1\}  \\
			\mu^s_{N+1} & = \begin{bmatrix} -(A'_{N+1} P_{N+1}+P_{N+1} A_{N+1})-(\hat{H}_{{N+1},{N+1}}+\hat{H}'_{{N+1},{N+1}}) + C_{N+1}'Q_{N+1} C_{N+1} & -P_{N+1} B_{N+1}^{(2)}+C'_{N+1} S_{N+1} \\
			-B_{N+1}^{(2)'}P_{N+1}+S'_{N+1} C_{N+1} & R_{N+1}\end{bmatrix}
			\\\mu^c_{N+1} &= \sum \limits_{j\in \mathcal{E}_{N+1}}\begin{bmatrix}\hat{H}'_{j,{N+1}}+\hat{H}_{{N+1},j} & 0\\ 0 & 0 \end{bmatrix} {\left(\mathcal{M}_{j}\right)}^{-1}\begin{bmatrix}\hat{H}'_{j,{N+1}}+\hat{H}_{{N+1},j} & 0\\ 0 & 0 \end{bmatrix}'
			\end{align}
		\end{subequations}
						\hrulefill
		%%%\vspace{-1em}
		% \setcounter{equation}{\value{MYtempeqncnt}}
%	}
\vspace{-0.5em}
\end{figure*}
\setcounter{equation}{13}
\vspace{0.5em}
\begin{corollary}
\label{cor:comp_synthesis}
Given a $QSR$-dissipative system $\mathbf{T_N}$ with energy matrices $P_i$ and messenger matrices $\mathcal{M}_i$, $i \in \mathbb{N}_N$ satisfying \eqref{seq_Cholesky} and \eqref{eq:messenger_matrix}, or \eqref{close_loop_seq_Cholesky} and \eqref{eq:messenger_matrix_control}, and a new subsystem $\Sigma_{N+1}$, the new networked system  $\mathbf{T_{N+1}}:=\mathbf{T_N}\vert \Sigma_{N+1}$ is $QSR$-dissipative with new
\begin{align*}
 Q &\mapsto \mbox{diag}(Q,Q_{N+1}), \quad Q_{N+1} \in \mathbb{R}^{m_{N+1} \times m_{N+1}}\\
 S &\mapsto \mbox{diag}(S,S_{N+1}), \quad S_{N+1}\in \mathbb{R}^{m_{N+1} \times l_{N+1}}\\
 R &\mapsto \mbox{diag}(R,R_{N+1}), \quad R_{N+1}\in \mathbb{R}^{l_{N+1} \times l_{N+1}},
\end{align*}
\noindent if there exist local control inputs \setcounter{equation}{14}
 \begin{equation}
      \begin{aligned}
     &u_{N+1}(t)=\sum \limits_{j \in \mathcal{E}_{N+1} \cup \{N+1\}}u_{N+1,j}(t),\\
     &u_{N+1,j}(t)= K_{N+1,j}x_j(t),  \\
     &u_{j}(t)\to u_j(t)+u_{j,N+1}(t),\\
     &u_{j,N+1}(t) = K_{j,N+1}x_{N+1}(t),
 \end{aligned}
 \end{equation}
\noindent $j \in \mathcal{E}_{N+1} \cup \{N+1\}$, such that 
\begin{align}\label{eq:compositional}
    \mathbb{P}_3:   \mbox{Find } \quad & P_{N+1}, K_{N+1,j}, K_{j,N+1}, K_{N+1,N+1} \, j \in \mathcal{E}_{N+1} \nonumber\\
       \mbox{s.t.} \quad & P_{N+1}>0,\nonumber\\ 
        & \mathcal{M}_{N+1}  > 0, \\
        & P_{N+1} \in \mathbb{R}^{n_{N+1} \times n_{N+1}}\nonumber
\end{align}
\noindent is feasible, where $\mathcal{M}_{N+1}$ is computed in \eqref{eq:messenger_matrix_comp_control}.
\end{corollary} 

Corollary \ref{cor:comp_synthesis} can be implemented algorithmically according to Algorithm \ref{algo:comp_synthesis}.

\begin{algorithm}[b]
	\caption{Compositional Synthesis for $\mathbf{T_{N+1}}{=}\mathbf{T_N}\vert\Sigma_{N+1}$}
	\label{algo:comp_synthesis}
	\hspace*{\algorithmicindent} \textbf{Given :} $\mathbf{T_N}$, $\Sigma_{N+1}$, $\mathcal{E}_{N+1}$, $\mathbf{T_{N+1}}=\mathbf{T_N}\vert\Sigma_{N+1}$, $\mathcal{M}_{i}$ and $P_i$, $i\in\mathbb{N}_{N}$\\
	\hspace*{\algorithmicindent} At subsystem $\Sigma_{N+1}:$
    % \hspace*{\algorithmicindent} \textbf{Output} 
	\begin{algorithmic}[1]
	   %  \State \Input .Initialize $i = 1$, $\mathcal{M}^{cl}_0 = 0$. %$\mathcal{M}_0 = 0$,
	   % \For{$i\leq \mathcal{E}_{N+1}$} \label{loop}
	    \State Receive $\mathcal{M}_{j}$ and $P_j$ from $\Sigma_{j}$, $j\in\mathcal{E}_{N+1}$.
	   % \State Set $\mathcal{M}_{i} = \mathcal{M}^{cl}_{i}$, $i\in\mathcal{E}_{N+1}$.
	    \If{$\mathbb{P}_1$ is feasible for $i=N+1$}
	    \State Compute $\mathcal{M}_{N+1}>0$ from \eqref{eq:messenger_matrix}. %and set $\mathcal{M}^{cl}_{N+1} = \mathcal{M}_{N+1}$.
	    \State Set $K_{N+1,N+1}=K_{N+1,j}=K_{j,N+1}=0$, $j\in\mathcal{E}_{N+1}$.
	    \Else
	    \State{\textit{Control design:}} Solve $\mathbb{P}_3$ to compute $K_{N+1,N+1}$, $K_{N+1,j}$, $K_{j,N+1}$, $j\in\mathcal{E}_{N+1}$, and $\mathcal{M}_{N+1}>0$, $P_{N+1}>0$ from  \eqref{eq:messenger_matrix_comp_control}.
	    \EndIf
	    \State Send $K_{j,N+1}$ to $\Sigma_{j}$, $j\in\mathcal{E}_{N+1}$.
% 		\State Set $i\mapsto i+1$.
% 		\EndWhile
	\end{algorithmic}
\end{algorithm}

In \cite{langbort2004distributed}, a similar dissipativity-based approach is used to derive LMIs for the synthesis of distributed controllers, wherein the coupling of the LMIs (and, thereby, the underlying controllers) follows the same structure as the interconnection graph of the network. These LMIs can be distributed numerically and solved iteratively at each subsystem using the method of alternating projections or subgradient methods \cite{langbort2004decomposition}\cite{langbort2003distributed}. In contrast, the synthesis LMIs in \eqref{eq:messenger_matrix_control} are distributed and are solved only once at each subsystem. Furthermore, we pose our synthesis as a sequential procedure, which allows for compositional synthesis for networks that may be expanded by adding new subsystems, as described in Section \ref{sec:seq}-C. When a new subsystem is added, the control synthesis problem $\mathbb{P}_3$ is only solved at the new subsystem, and is independent of the size of the existing networked system.

\section{Numerical Example}\label{sec:Example}
\begin{figure}[b]
	\centering
	\vspace{1.3em}
	\includegraphics[scale=0.7,trim=0.1cm 0cm 0cm 0cm]{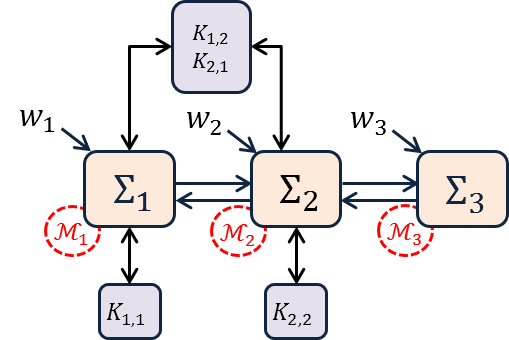}
	\caption{Schematic of the networked dynamical system $\mathbf{T_3}$ with control architecture.}
	\label{fig:T3}
	%%\vspace{-1em}
	% \hrulefill
\end{figure} 
\begin{figure*}[b]
	\centering
	\includegraphics[scale=0.75]{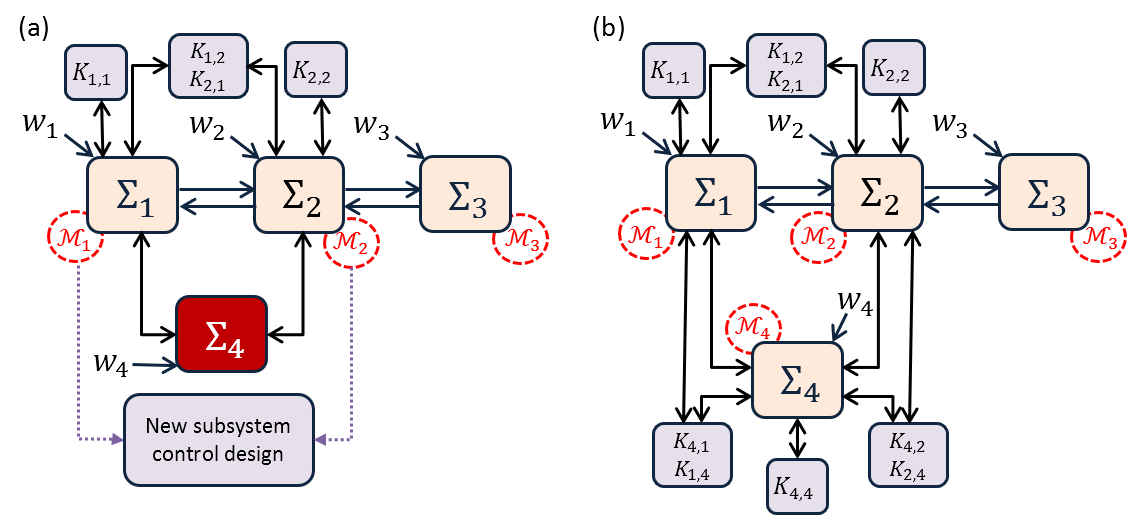}
	\caption{(a) Schematic of compositional control design for $\mathbf{T_{4}}=\mathbf{T_3}\vert\Sigma_{4}$ when a new subsystem $\Sigma_{4}$ is connected to $\mathbf{T_3}$, (b) Networked dynamical system $\mathbf{T_4}$ with control architecture.}
	\label{fig:T4}
	% %%\vspace{-0.1em}
	% \hrulefill
\end{figure*}
In this section, we present a numerical example to illustrate the distributed synthesis of local controllers for networked dynamical systems, and demonstrate the compositionality of the approach when new subsystems are added to the existing network. We begin by considering a
networked system $\mathbf{T_3}$, comprised of three subsystems with dynamics and coupling given by

\setcounter{equation}{17}
\begin{align}
&\Sigma_1: & \dot{x}_1(t) & = \begin{bmatrix} -9 & 1\\ 5 & 7\end{bmatrix} x_1 (t)\! +\!  v_1(t)\! +\! \begin{bmatrix} 1 \\ 0.5 \end{bmatrix} w_1(t)\! + u_1(t) \nonumber \\
& &y_1(t) & = \begin{bmatrix} 3 & 2 \end{bmatrix} x_1(t) \\
& &v_1(t) & = \begin{bmatrix} 0.5 & -0.7 \end{bmatrix} x_1(t) + 0.1 x_2(t) %\nonumber.\\
\end{align}
\begin{align}
&\Sigma_2: &\dot{x}_2(t) & = 3 x_2 (t)\! +\! v_2(t)\! +\!  w_2(t)\! +\! u_2(t) \nonumber \\
& &y_2(t) & = x_2(t) \\
& &v_2(t) & = \begin{bmatrix} 1 & -0.5 \end{bmatrix} x_1(t) +  0.5 x_2(t) -0.1 x_3(t) \nonumber.\\
&\Sigma_3: &\dot{x}_3(t) & = - x_3 (t)\! +\!  v_3(t)\! +\!  w_3(t)\! +\!  u_3(t) \nonumber \\
& &y_3(t) & = x_3(t) \\
& &v_3(t) & = -0.7 x_2(t) + 0.2 x_3(t) \nonumber.
\end{align}

The objective is to guarantee passivity of the networked system according to the definition in Remark \ref{rem:cases}-(1). 

We begin by checking if $\Sigma_1$ is passive using Algorithm \ref{algo:seq_synthesis}, and compute  controller matrix $K_{1,1}$ to guarantee passivity of $\Sigma_1$. We also compute the closed loop messenger and energy matrices $\mathcal{M}_1$ and $P_1$ respectively of $\Sigma_1$. We then use $\mathcal{M}_1$ and $P_1$ communicated from $\Sigma_1$, and the dynamics of $\Sigma_2$ to 

\noindent verify the sufficient conditions in Theorem \ref{thm:seq_analysis} for the network comprised of $\Sigma_1$ and $\Sigma_2$. Since the sufficient conditions are not satisfied, we use the procedure in Algorithm \ref{algo:seq_synthesis} to synthesize controller matrices $K_{2,2}$, $K_{2,1}$ and $K_{1,2}$ at $\Sigma_2$ to guarantee passivity of the interconnection of $\Sigma_1$ and $\Sigma_2$. Additionally, we compute messenger matrix $\mathcal{M}_2$ and energy matrix $P_2$ at $\Sigma_2$. Next, we use the dynamics of the subsystem $\Sigma_3$, and $\mathcal{M}_2$ and $P_2$ communicated from $\Sigma_2$ to $\Sigma_3$ in Algorithm \ref{algo:seq_synthesis}. Since $\mathbb{P}_1$ is feasible (Step 6 in Algorithm \ref{algo:seq_synthesis}), the networked system $\mathbf{T_3}$ comprised of the interconnection of $\Sigma_3$ with $\Sigma_1$ and $\Sigma_2$ is passive, and no controller design is required at $\Sigma_3$. 

\begin{figure}[H]
	\vspace{0.5em}
	\centering
	\frame{\includegraphics[scale=0.8]{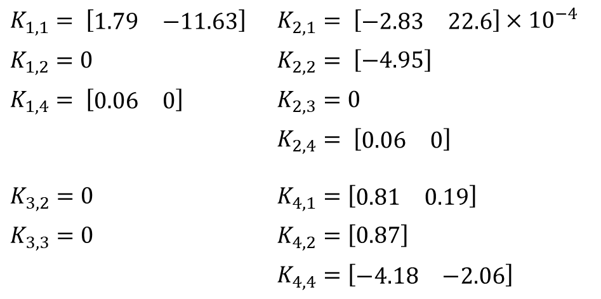}}
	\caption{Controller matrices for networked system $\mathbf{T_{4}}$ with solution sequence \{1-2-3-4\}.}
	\label{fig:controllers}
	\vspace{-0.5em}
	% \hrulefill
\end{figure}

The controller matrices $K_{3,3}$, $K_{2,3}$ and $K_{3,2}$ are set to zero, and messenger matrix $\mathcal{M}_3$ and energy matrix $P_3$ are computed and stored at $\Sigma_3$. The networked system $\mathbf{T_3}$ with its control architecture is shown in Fig. \ref{fig:T3}, and the controller matrices are as shown in Fig. \ref{fig:controllers}.

Now consider the networked system $\mathbf{T_4}=\mathbf{T_3}|\Sigma_4$ formed by adding a new subsystem $\Sigma_4$ to $\mathbf{T_3}$ as shown in Fig. \ref{fig:T4}(a), where $\Sigma_4$ is dynamically coupled to $\Sigma_1$ and $\Sigma_2$. The dynamics of $\Sigma_4$ is given by

\begin{align}
\Sigma_4: \dot{x}_4(t) & \! =\!  \begin{bmatrix} 2 & 1\\ 3 & 0.8\end{bmatrix} x_4(t)\! +\! \begin{bmatrix} 1.2 \\ 0.8 \end{bmatrix} v_1(t)\! +\! \begin{bmatrix} 0.5 \\ -0.2 \end{bmatrix} w_4(t)\! \nonumber \\ & \ +\! \begin{bmatrix} 1.2 \\ 0.8 \end{bmatrix} u_4(t) \nonumber \\
y_4(t) & \! =\!  \begin{bmatrix} 2.1 & 0.6 \end{bmatrix} x_4(t) \\
v_4(t) & \! =\!  \begin{bmatrix}-0.9& -0.3 \end{bmatrix}\! x_1(t)\! -\! 0.9 x_2(t)\! +\! \begin{bmatrix} 1.1 & 0.4 \end{bmatrix} \! x_4(t).\nonumber
\end{align} \normalsize

Additionally,~the~coupling~inputs~$v_1(t)$~and~$v_2(t)$~are~updated~to,
\begin{align*}
v_1(t) & \!= \!\begin{bmatrix} 0.5 & -0.7 \end{bmatrix}\! x_1(t) \!+\! 0.1 x_2(t)\!+\!\begin{bmatrix}0.2 & 0.2\end{bmatrix}\!x_4(t), \\
v_2(t) &\! = \!\begin{bmatrix} 1 & -0.5 \end{bmatrix} \!x_1(t) \!+\!  0.5\! x_2(t) \!-\!0.1 \!x_3(t)\!+\!\begin{bmatrix}0.2 & 0.2\end{bmatrix}\!x_4(t).
\end{align*}
At subsystem $\Sigma_4$, we use matrices $\mathcal{M}_1$, $P_1$, $\mathcal{M}_2$ and $P_2$ received from $\Sigma_1$ and $\Sigma_2$ (its neighboring subsystems) in the compositional synthesis procedure described in Algorithm \ref{algo:comp_synthesis} to design controller matrices $K_{4,4}$, $K_{1,4}$, $K_{4,1}$, $K_{4,2}$ and $K_{2,4}$ that guarantee passivity of the networked system $\mathbf{T_4}=\mathbf{T_3}|\Sigma_4$. The compositional control synthesis procedure is illustrated in Fig. \ref{fig:T4}. The distributed synthesis algorithm allows for dynamics of subsystems to be dissimilar and of different dimensions, as long as the network is `proper', that is, the input-output dimensions are suitable to define the interconnections between subsystems.

%The distributed synthesis procedure does not impose weak coupling between subsystems, as evident from sparsity structure of $H+K$, as shown in Fig. \ref{}.

\begin{figure}[t]
	%%\vspace{-1em}
	\centering
	\includegraphics[scale=0.6]{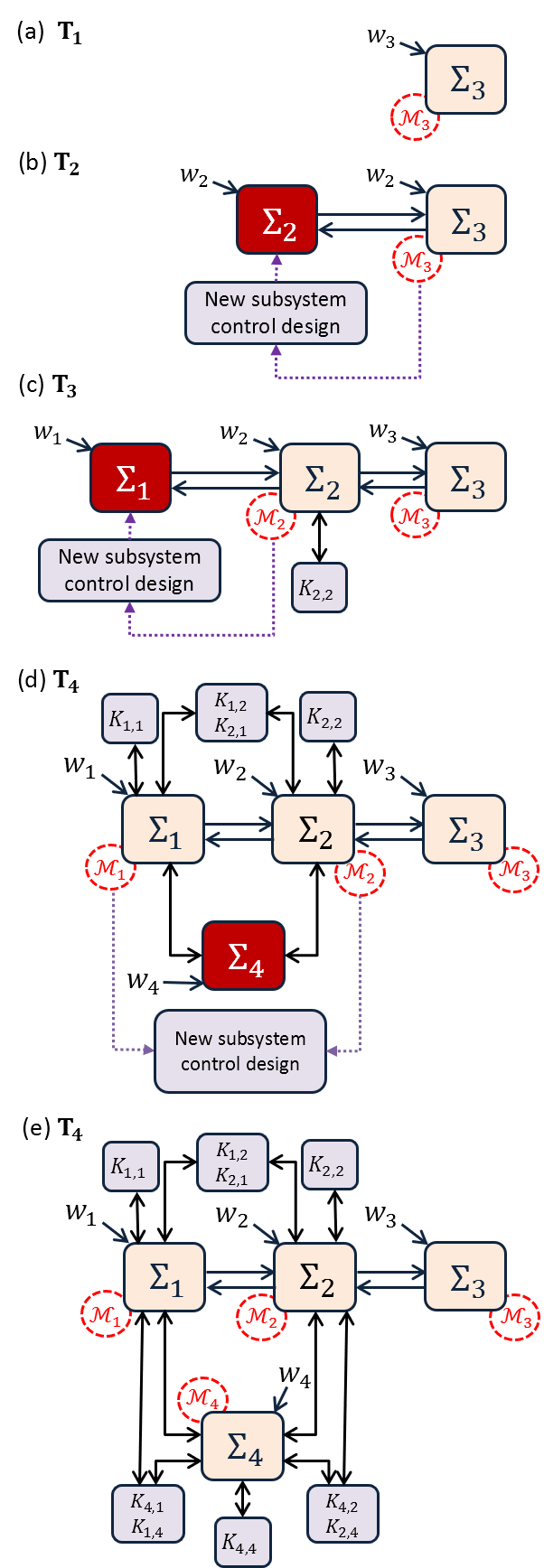}
	\caption{Schematic of distributed control synthesis for networked system $\mathbf{T_{4}}$ with solution sequence \{3-2-1-4\}.}
	\label{fig:example_3214}
	\vspace{-1.2em}
	% \hrulefill
\end{figure}

We would like to highlight the fact that the proposed algorithm distributes the analysis and control synthesis between subsystems  \textit{sequentially}. As described in Remark \ref{rem:feasibility}, Algorithm \ref{algo:seq_synthesis} is generally always feasible (with only one exception). In the next part of this example, we demonstrate the feasibility of the control synthesis by reapplying the distributed synthesis approach to the same network, but solving the synthesis problem in Algorithm \ref{algo:seq_synthesis} in two different sequences as shown in Fig. \ref{fig:example_3214} and Fig. \ref{fig:example_3412}.

\subsubsection{Solution sequence $\{3-2-1-4\}$}
Consider the set-up shown in Fig. \ref{fig:example_3214}. As opposed to the discussion so far in this example, where the distributed synthesis Algorithm \ref{algo:seq_synthesis} solves $\mathbb{P}_1$ (and $\mathbb{P}_2$) for $\Sigma_1$ first, followed by $\Sigma_2$, $\Sigma_3$, and $\Sigma_4$, in that order, we now
begin by analyzing $\Sigma_3$ first, followed by $\Sigma_2$, $\Sigma_1$, and $\Sigma_4$, in that order. The design steps are summarized below:
\begin{itemize}
    \item[(a)] We use the sufficient conditions in Theorem \ref{thm:seq_analysis} to verify that $\Sigma_3$ is passive. Therefore, no controller design is required at $\Sigma_3$, and we set $K_{3,3}=0$. We also compute and store the messenger and energy matrices $\mathcal{M}_3$ and $P_3$ respectively of $\Sigma_3$. 
    \item[(b)] We then consider the network $\mathbf{T}_2$, comprised of the interconnection of $\Sigma_3$ and $\Sigma_2$, as shown in Fig. \ref{fig:example_3214}-(b). We use the messenger and energy matrices $\mathcal{M}_3$ and $P_3$ respectively, communicated from $\Sigma_3$, along with the dynamics of $\Sigma_2$ in Algorithm \ref{algo:seq_synthesis} to verify the passivity of this network. Since the passivity conditions for $\mathbf{T}_2$ are not met, we use Step 10 of Algorithm \ref{algo:seq_synthesis} to synthesize controller matrices $K_{2,2}$, $K_{2,3}$, and $K_{3,2}$ at $\Sigma_2$ to guarantee passivity of interconnection of $\Sigma_3$ and $\Sigma_2$. In this case, the controller matrices $K_{2,3}$, and $K_{3,2}$ turn out to be zero. We also compute messenger matrix $\mathcal{M}_2$ and energy matrix $P_2$ at $\Sigma_2$.
    \item[(c)] Next, we consider the network $\mathbf{T}_3=\mathbf{T}_2|\Sigma_1$, as shown in Fig. \ref{fig:example_3214}-(c). We use the dynamics of the subsystem $\Sigma_1$, and $\mathcal{M}_2$ and $P_2$ communicated from $\Sigma_2$ to $\Sigma_1$ in Algorithm \ref{algo:seq_synthesis} to verify that the network $\mathbf{T}_3$ is not passive, that is, problem $\mathbb{P}_1$ is infeasible. Therefore, we solve $\mathbb{P}_2$ to design controller matrices 
    $K_{1,1}$, $K_{1,2}$, and $K_{2,1}$ at $\Sigma_1$ to guarantee passivity of the networked system $\mathbf{T_3}$ comprised of the interconnection of $\Sigma_3$, $\Sigma_1$ and $\Sigma_2$. We also compute messenger matrix $\mathcal{M}_1$ and energy matrix $P_1$ at $\Sigma_1$.
    \item[(d)] Lastly, we apply Algorithm \ref{algo:seq_synthesis} to subsystem $\Sigma_4$. The dynamics of the subsystem $\Sigma_4$, and messenger matrices $\mathcal{M}_1$ and $\mathcal{M}_2$, and energy matrices $P_1$ and $P_2$, communicated to $\Sigma_4$ from its neighboring subsystems $\Sigma_1$ and $\Sigma_2$, are used in $\mathbb{P}_1$ and $\mathbb{P}_2$ to solve for the controller gains $K_{4,4}$, $K_{4,1}$, $K_{4,2}$, $K_{1,4}$, and $K_{2,4}$. The closed loop system $\mathbf{T_4}$ is guaranteed to be passive.  
\end{itemize}

The designed controller matrices are shown in Fig. \ref{fig:controllers_3214}. Note that although the distributed synthesis algorithm was applied in the sequence $\{3-2-1-4\}$, the synthesis problem is feasible, and the designed local controllers look very similar to the ones designed when the synthesis algorithm was solved in the sequence $\{1-2-3-4\}$.
\subsubsection{Solution sequence $\{3-4-1-2\}$}
A more interesting scenario is represented in Fig. \ref{fig:example_3412}, where the sequence in which Algorithm \ref{fig:example_3412} is applied starts from $\Sigma_3$, and moves on to $\Sigma_4$ in the next step, to guarantee passivity of the network $\mathbf{T_2}$ comprised of $\Sigma_3$ and $\Sigma_4$.  Since the two subsystems, $\Sigma_3$ and $\Sigma_4$, have no direct coupling, $\mathbf{T}_2$ is not fully connected. 
\begin{figure}[t]
	\vspace{0.5em}
	\centering
	\frame{\includegraphics[scale=0.8]{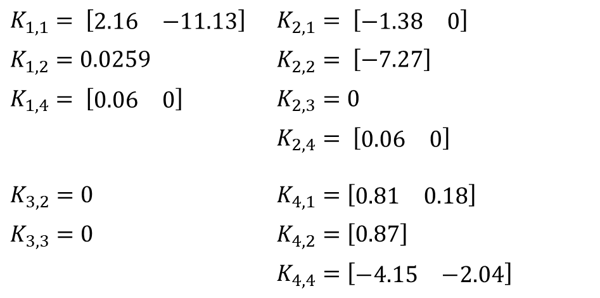}}
	\caption{Controller matrices for networked system $\mathbf{T_{4}}$ with solution sequence \{3-2-1-4\}.}
	\label{fig:controllers_3214}
	% %%\vspace{-0.1em}
	% \hrulefill
\end{figure}

However, the results presented in Theorem \ref{thm:seq_analysis} and Theorem \ref{thm:distributed_synthesis} do not impose any restrictions on the topology of the network and can be applied regardless of the connectivity of the graph. Therefore, when the distributed synthesis algorithm is applied to $\Sigma_4$ here, it uses the dynamics of $\Sigma_4$ in $\mathbb{P}_1$ and $\mathbb{P}_2$ to compute controller gain $K_{4,4}$, and messenger matrix $\mathcal{M}_4$ and energy matrix $P_4$. The messenger and energy matrices of $\Sigma_3$ are \textit{not} communicated to $\Sigma_4$ because $\Sigma_3$ and $\Sigma_4$ are not coupled. In the next step, $\mathcal{M}_4$ and $P_4$ are communicated to $\Sigma_1$ from $\Sigma_4$ and Algorithm \ref{algo:seq_synthesis} is applied to $\Sigma_1$. Note that at this stage, the network $\mathbf{T_3}$ comprised of $\Sigma_1$, $\Sigma_3$, and $\Sigma_4$ is still not fully connected. However, the proposed distributed synthesis approach still works to guarantee the passivity of the network $\mathbf{T_3}$ until this point. Similarly, $\Sigma_2$ is added to the network and Algorithm \ref{algo:seq_synthesis} is applied to $\Sigma_2$, using its dynamics, and the messenger and energy matrices, $\mathcal{M}_1$, $\mathcal{M}_3$, $\mathcal{M}_4$, $\mathcal{P}_1$, $\mathcal{P}_3$ and $\mathcal{P}_4$, received from its neighbors $\Sigma_1$, $\Sigma_3$ and $\Sigma_4$, to guarantee passivity of the networked system $\mathbf{T}_4$. The steps involved in the synthesis are illustrated in Fig. \ref{fig:example_3412}, and the designed controller matrices are shown in Fig. \ref{fig:controllers_3412}.

\begin{remark}
Note that the controller gains designed at $\Sigma_2$ using this solution sequence are quite different from the ones designed in Fig. \ref{fig:controllers} and Fig. \ref{fig:controllers_3214}. This difference can be interpreted in the context of Remark \ref{rem:messengermatrix} as follows. For the solution sequence $\{1-2-3-4\}$, when the control design process is
\begin{figure}[H]
	%%\vspace{-1em}
	\centering
	\includegraphics[scale=0.628]{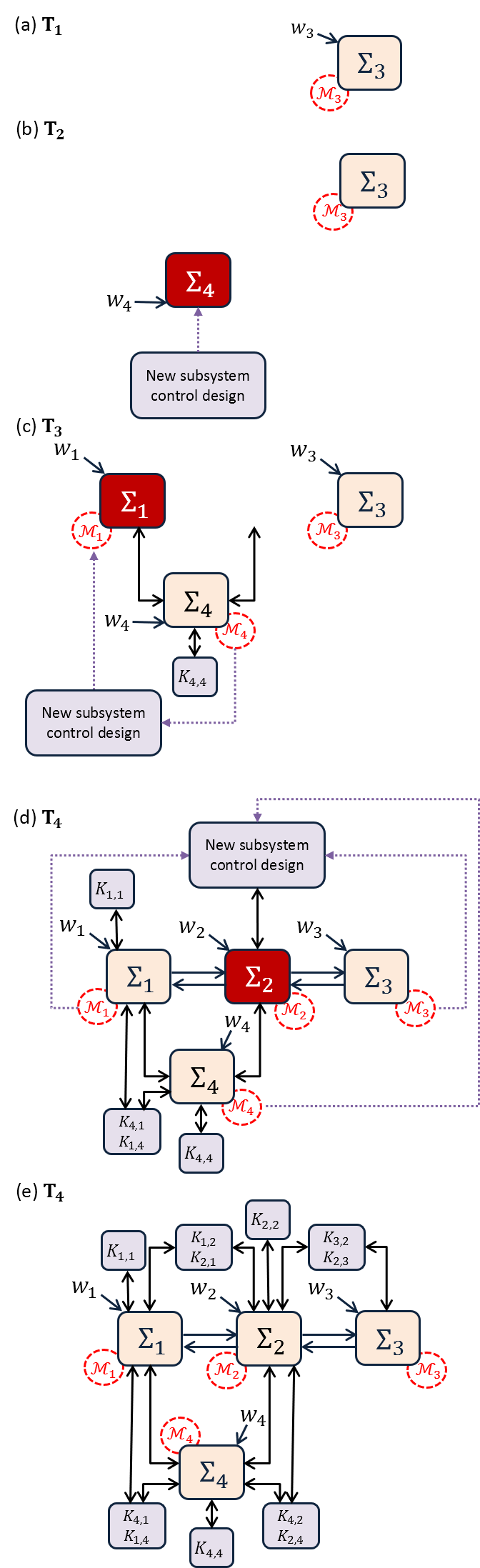}
	\caption{Schematic of distributed control synthesis for networked system $\mathbf{T_{4}}$ with solution sequence \{3-4-1-2\}.}
	\label{fig:example_3412}
	% %%\vspace{-0.1em}
	% \hrulefill
\end{figure}
\noindent carried out at $\Sigma_2$, the term $\mu_2^c$ in the messenger matrix $\mathcal{M}_2$ corresponding to the energy flow from its neighbors only comprises of the energy flow from subsystem $\Sigma_1$. Similarly, for the solution sequence $\{3-2-1-4\}$, the messenger matrix $\mathcal{M}_2$ only contains the energy flow from the neighbor $\Sigma_3$. However, when the control synthesis algorithm is solved in the sequence $\{3-4-1-2\}$, the subsystem $\Sigma_2$ is added in the last step of the procedure, and the term $\mu_2^c$ in the messenger matrix $\mathcal{M}_2$ comprises of energy flows from all its neighboring subsystems, $\Sigma_1$, $\Sigma_2$, and $\Sigma_3$. This results in large gains for the controllers designed at $\Sigma_2$ to ensure the positive definiteness of the messenger matrix $\mathcal{M}_2$. This example highlights the trade-off involved in the proposed distributed synthesis, introduced by the sequential nature of the synthesis algorithm. %If the dynamics of all the subsystems were known, then, the sequence in which the synthesis is carried out can conceivably be assigned based on the dissipativity levels of the subsystems and the energy flows between subsystems. However, in that case, the synthesis process will no longer be 
%truly distributed; the resulting synthesis will be hierarchical, where the determination of the sequence is centralized, and the computation of the controllers is local to the subsystems. 
In achieving distributed synthesis, Algorithm \ref{algo:seq_synthesis} introduces some conservatism, where local controllers may be designed at an intermediate step of the synthesis procedure, even if the networked system as a whole is dissipative. Intuitively, the design of redundant or unnecessary controllers will decrease with the number of subsystems into which the networked system is divided, with a purely centralized synthesis being the least conservative in this regard. However, we note that it is precisely this sequential nature of the synthesis that allows compositionality for networks that may be expanded by adding new subsystems, as described in Section \ref{sec:seq}-C.
\end{remark}

\begin{figure}[t]
	\vspace{0.5em}
	\centering
	\frame{\includegraphics[scale=0.8]{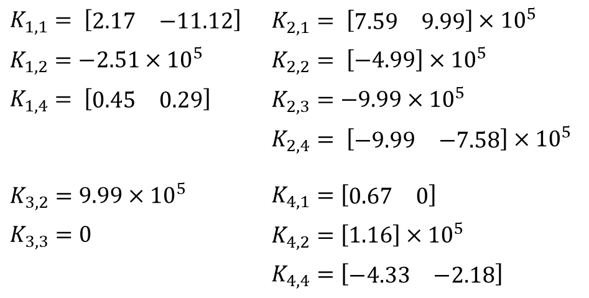}}
	\caption{Controller matrices for networked system $\mathbf{T_{4}}$ with solution sequence \{3-4-1-2\}.}
	\label{fig:controllers_3412}
	% %%\vspace{-0.1em}
	% \hrulefill
\end{figure}

\begin{figure*}[b]
	\setcounter{equation}{26}
		\hrulefill
	%%\vspace{0.5em}
	\begin{subequations}\label{eq:messenger_matrix_control_switched}
		\begin{align}
		\mathcal{\overline M}_i &= \begin{cases} \mu^s_{1}, & i =1  \\ \mu^s_{i} - \mu^c_{i}, & i \in \mathbb{N}_N \backslash \{1\}
		\end{cases} \label{eq:mi_switched} \\
		% Q & = diag (Q_1, Q_2, \ldots, Q_N) \\
		\hat{H}_{i,j}^{\sigma_i} &= P_i(B_i^{(1)^{\sigma_i}}H_{i,j}+B_i^{(3)^{\sigma_i}}K_{i,j}), \hspace{20.2em} i\in\mathbb{N}_N,\, j \in \mathcal{E}_i \\
		\mu^s_{i} & = \begin{bmatrix} -(A_i^{\sigma_i'} P_i+P_i A_i^{\sigma_i'})-(\hat{H}_{i,i}^{\sigma_i}+\hat{H}_{i,i}^{\sigma_i'}) + C_i^{\sigma_i'}Q_i C_i^{\sigma_i} & -P_i B_i^{(2)^{\sigma_i}}+C_i^{\sigma_i'} S_i \\
		-B_i^{(2)^{\sigma_i'}}P_i+S'_i C_i^{\sigma_i} & R_i\end{bmatrix}, \hspace{1.3em} i \in \mathbb{N}_N
		\\\mu^c_{i} &= \sum \limits_{j\in \mathcal{E}_i \cap \mathbb{N}_{i-1}}  \begin{bmatrix}\hat{H}_{j,i}^{\sigma_j'}+\hat{H}_{i,j}^{\sigma_i} & 0\\ 0 & 0 \end{bmatrix} \left(\mathcal{M}_{j}\right)^{-1}\begin{bmatrix}\hat{H}_{j,i}^{\sigma_j'}+\hat{H}_{i,j}^{\sigma_i} & 0\\ 0 & 0 \end{bmatrix}', \hspace{8.7em} i \in \mathbb{N}_N\backslash \{1\} \label{eq:mu_cs} \\
		\mathcal{M}_i & =\arg\min\limits_{{\sigma_i,\sigma_j}} ||\mathcal{\overline M}_i||, \hspace{25.7em} i \in \mathbb{N}_N
		\end{align}
	\end{subequations}
	%	%%\vspace{-1em}
\end{figure*}

\section{Extension to Switched Systems}\label{sec:switched}
In many emerging applications of large-scale networked systems in infrastructure networks, the subsystem dynamics, and even the coupling matrices between subsystems can change during operation. For example, in power grids comprised of interconnected microgrids, both the dynamics of individual microgrids and the coupling between microgrids change when a new microgrid is connected to the network, and with changes in the operating point \cite{agarwal2017feedback}. In order to synthesize local controllers in a distributed manner and guarantee compositionality for such applications, we extend the distributed synthesis presented in Section \ref{sec:seq} to networks of switched systems, where subsystem dynamics is time-varying. 
%\begin{equation}\label{switched system}
%\begin{aligned}
%\dot{x}_i(t) & = A_{\sigma_i(t)}^i x_i(t) + B_{\sigma_i(t)}^{i^{(1)}}v_i(t) + B_{\sigma_i(t)}^{i^{(2)}} w_i(t) + B_{\sigma_i(t)}^{i^{(3)}} u_i(t), \\
%y_i(t) & = C_{\sigma_i(t)}^i x_i(t), \\
%v_i(t) & = \sum \limits_{j \in \mathbb{N}_N} H_{i,j} x_{j}(t),
%\end{aligned}
%\end{equation}

Consider a networked system $\mathbf{T_N^{\sigma}}$ comprised of $N$ subsystems, where the dynamics of the $i$-th subsystem $\Sigma_{i}$ is switching and is given by

\setcounter{equation}{22}
\begin{equation}\label{switched system}
\begin{aligned}
\dot{x}_i(t) & = A_{i}^{\sigma_i(t)} x_i(t) + B_{i}^{(1)^{\sigma_i(t)}}v_i(t) + B_{i}^{(2)^{\sigma_i(t)}} w_i(t) \\ & \quad + B_{i}^{(3)^{\sigma_i(t)}} u_i(t), \\
y_i(t) & = C_{i}^{\sigma_i(t)} x_i(t), \\
v_i(t) & = \sum \limits_{j \in \mathbb{N}_N} H_{i,j} x_{j}(t),
\end{aligned}
\end{equation}

\noindent where \(x_i(t)\), \(y_i(t)\), \(v_i(t)\), \(u_i(t)\), and \(w_i(t)\) are as described in Section \ref{sec:prob}. The system matrices  $A_{i}^{\sigma_i(t)}$, $B_{i}^{(1)^{\sigma_i(t)}}$, $B_{i}^{(2)^{\sigma_i(t)}}$, $B_{i}^{(3)^{\sigma_i(t)}}$, and $C_{i}^{\sigma_i(t)}$ vary based on the value of the switching signal $\sigma_{i}(t): \mathbb{R}^{+}\to \mathbb{N}_{\eta_i}$, where $\eta_i$ is the number of switching modes of $\Sigma_{i}$. Note that we do not place any restrictions on the sequence in which the dynamics of $\Sigma_{i}$ switches, and do not require the switching sequence to be known a priori. 

Now, the dynamics of $\mathbf{T_N^{\sigma}}$ is described by

\begin{equation}\label{interconnected_switched}
\begin{aligned}
\dot{x}(t) & = A^{\sigma(t)}x(t) + B^{(1)^{\sigma(t)}}v(t) + B^{(2)^{\sigma(t)}}w(t) \\
& \quad + B^{(3)^{\sigma(t)}}u(t)\\
y(t) & = C^{\sigma(t)}x(t)\\
v(t) & = Hx(t)
\end{aligned}
\end{equation}

\noindent where
\begin{align*}
A^{\sigma(t)} &=\mbox{diag}(A_1^{\sigma_1(t)}, A_2^{\sigma_2(t)}, \dots, A_N^{\sigma_N(t)})\\
B^{(j)^{\sigma(t)}} &= \mbox{diag}(B_1^{(j)^{\sigma_1(t)}}, B_2^{(j)^{\sigma_2(t)}}, \dots,B_N^{(j)^{\sigma_N(t)}}), \quad j\in \mathbb{N}_3\\
C^{\sigma(t)}  &=\mbox{diag}(C_1^{\sigma_1(t)}, C_2^{\sigma_2(t)}, \dots, C_N^{\sigma_N(t)})\\
H  &=  \begin{bmatrix} \nonumber 
H_{i,j}
\end{bmatrix}_{i,j \in \mathbb{N}_N}
\end{align*}

\noindent and $x(t)$, $y(t)$, $v(t)$, $u(t)$, $w(t)$ and $\sigma(t)$ are the augmented system state, output, coupling input, control input, disturbance and switching signal formed by stacking $x_{i}(t)$, $y_{i}(t)$, $v_{i}(t)$, $u_{i}(t)$, $w_{i}(t)$ and $\sigma_{i}(t)$ respectively of all $N$ subsystems. 

As described in \cite{zhao2008dissipativity}, the classical form of dissipativity in Definition \ref{def:diss} holds for switched system \eqref{interconnected_switched} as well. Along the lines of Section \ref{sec:seq}, we have the following result on distributed synthesis of local controllers to guarantee dissipativity of the networked switched system $\mathbf{T_N^{\sigma}}$. 

\begin{theorem}\label{thm:distributed_synthesis_switched}
	The local control inputs 
	
	\begin{equation}
	\begin{aligned}
	&u_i(t)=\sum \limits_{j \in \mathcal{E}_i \cup \{i\}}u_{i,j}(t), \;  i \in \mathbb{N}_N,\\
	&u_{i,j}(t)= K_{i,j}x_j(t), \; j \in \mathcal{E}_i \cup \{i\},
	\end{aligned}
	\end{equation}
	
\noindent	designed by solving %\setcounter{equation}{23}

	\begin{equation}\label{close_loop_seq_Cholesky_switched}
	\begin{aligned}
	\mathbb{P}_4:   \mbox{Find } \quad & P_i, K_{i,i}, K_{i,j}, K_{j,i} \, j \in \mathcal{E}_i \\
	\mbox{s.t.} \quad & P_i>0,\\
	& {\mathcal{\overline{M}}}_i  > 0, \\
	% & \left [ \begin{array}{c;{2pt/2pt}c}
	% \mu_i^s & \begin{array}{c;{2pt/2pt}c}
	% \end{array} \\ \hdashline[2pt/2pt]
	% -B^{(2)'}P+S'C & R
	% \end{array}\right ]
	% \begin{bmatrix}
	% \mu_i^s & 
	% \end{bmatrix}
	& P_i \in \mathbb{R}^{n_i \times n_i}, \\
	&K_{i,i} \in \mathbb{R}^{p_i \times n_i}, \\
	& K_{i,j} \in \mathbb{R}^{p_i \times n_j}, \, K_{j,i} \in \mathbb{R}^{p_j \times n_i},
	\end{aligned}
	\end{equation}
	
\noindent	for all $i\in\mathbb{N}_N$ and all $\sigma_i \in \mathbb{N}_{\eta_i}$, $\sigma_j \in \mathbb{N}_{\eta_j}$, $j \in \mathcal{E}_i \cap \mathbb{N}_{i-1}$, where $\mathcal{\overline M}_i$ is computed from \eqref{eq:messenger_matrix_control_switched}, render $\mathbf{T_N^{\sigma}}$ \eqref{interconnected_switched} $QSR$-dissipative with
\begin{align*}
Q &= \mbox{diag}(Q_1, Q_2, \dots, Q_N), \quad Q_i~\in~\mathbb{R}^{m_i \times m_i}\\
S &= \mbox{diag}(S_1, S_2, \dots, S_N), \quad S_i \in \mathbb{R}^{m_i \times l_i}\\
R &= \mbox{diag}(R_1, R_2, \dots, R_N), \quad R_i\in \mathbb{R}^{l_i \times l_i}, \quad i \in \mathbb{N}_N.\\
\vspace{-3em}
\end{align*}
The closed-loop messenger matrix of $\Sigma_i$ is then given by $\mathcal{M}_i=\arg\min\limits_{{\sigma_i,\sigma_{j}}} ||\mathcal{\overline M}_i|| $.
\end{theorem}
\vspace{0.5em}

The messenger matrix $\mathcal{M}_i$ in the distributed synthesis result of Theorem \ref{thm:distributed_synthesis_switched} corresponds to the least dissipative mode of $\Sigma_i$. This allows for a reduction in the computational complexity of the control synthesis arising from a possibly large number of switching modes in the networked system. If the coupling matrix $H$ is also switching, then the maximum value of the coupling term $\mu_{i,\sigma_i}^c$ in \eqref{eq:mu_cs} over all possible switching sequences of $\Sigma_j, j\in \mathcal{E}_i \cap \mathbb{N}_{i-1}$ can be used in \eqref{eq:mi_switched}.

%We make the following remarks about the distributed synthesis of local controllers for networks of switched systems in Theorem \ref{thm:distributed_synthesis_switched}.
%\begin{itemize}
%	\item Centralized control design for a networked switched system $\mathbf{T_N^{\sigma}}$ will requin Theorem \ref{thm:distributed_synthesis_switched} requires the 
%\end{itemize}
\vspace{0.5em}
\begin{remark}\label{rem:lyapunov} Note that Theorem \ref{thm:distributed_synthesis_switched} is based on using the same energy matrix $P_i$ for all switching modes of $\Sigma_i$ in \eqref{switched system}. We can allow different energy matrices in different switching modes of the subsystem for specific switching signals that are known a priori. However, it can be shown that a dissipative switched system under arbitrary switching can not have different energy matrices in different switching modes (see Appendix for details).\end{remark}
\vspace{0.5em}
\begin{figure*}[b]
	\setcounter{equation}{27}
	%%\vspace{0.5em}
	\hrulefill
	\begin{subequations}\label{eq:microgrid_dynamics}
		\begin{align}
		A_i^{\sigma_i} &= \begin{bmatrix}
		-\frac{1}{X_{C,ti}}\left(\sum\limits_{j\in\mathcal{S}_{i,\sigma_i}} \frac{X_{R,ij}}{Z_{ij}^2} \right) & \omega_0 - \frac{1}{X_{C,ti}}\left(\sum\limits_{j\in\mathcal{S}_{i,\sigma_i}} \frac{\omega_0 X_{L,ij}}{Z_{ij}^2} \right) & \frac{k_i}{X_{C,ti}} & 0 \label{eq:microgrid_dynamics_switching} \\
		-\omega_0 + \frac{1}{X_{C,ti}}\left(\sum\limits_{j\in\mathcal{S}_{i,\sigma_i}} \frac{\omega_0 X_{L,ij}}{Z_{ij}^2} \right) & -\frac{1}{X_{C,ti}}\left(\sum\limits_{j\in\mathcal{S}_{i,\sigma_i}} \frac{X_{R,ij}}{Z_{ij}^2} \right) & 0 & \frac{k_i}{X_{C,ti}} \\
		-\frac{k_i}{X_{L,ti}} & 0 & \frac{X_{R,ti}}{X_{L,ti}} & \omega_0 \\
		0 & -\frac{k_i}{X_{L,ti}} & -\omega_0 & -\frac{X_{R,ti}}{X_{L,ti}}
		\end{bmatrix}, \quad \begin{matrix*}[l]
		\mathcal{S}_{1,\sigma_1} = \begin{cases}
		\{2\}, &\sigma_1 = 1\\
		\{2,3\}, &\sigma_1 = 2
		\end{cases}\\
		\mathcal{S}_{2,\sigma_2}  = \{1\}, \hspace{2.6em}\sigma_2 = 1 \\
		\mathcal{S}_{3,\sigma_3}  = \{1\}, \hspace{2.6em}\sigma_3 = 1
		\end{matrix*}\\ 
		%	A_{i_{2,1}}  = - \begin{bmatrix}
		%	1 & 0 & 0 & 0 \\
		%	0 & 1 & 0 & 0
		%	\end{bmatrix} \\
		%	A_i^{\sigma_i} & =  \left[\begin{array}{c;{2pt/2pt}c}
		%	A_{i_{1,1}}^{\sigma_i}  & 0 \\ \hdashline[2pt/2pt]
		%	 A_{i_{2,1}} & 0
		%	\end{array}\right],	\quad
		B_{i}^{(1)^{\sigma_i}} &= \mathbf{I}, \quad
		B_{i}^{(2)^{\sigma_i}} = \begin{bmatrix}
		-\frac{1}{X_{C,ti}} & 0 & 0 & 0   \\
		0 & -\frac{1}{X_{C,ti}} & 0 & 0 
		\end{bmatrix}', \quad
		B_{i}^{(3)^{\sigma_i}} =\begin{bmatrix}
		0 & 0 & 	\frac{1}{X_{L,ti}} & 0 \\
		0 & 0 & 0 &  \frac{1}{X_{L,ti}}
		\end{bmatrix}' \\
		% 6 by 6 matrix
		C_i^{\sigma_i}  &= \mathbf{I}, \quad H_{i,j_{1,1}} = \frac{1}{X_{C,ti}} \begin{bmatrix}
		\frac{X_{R,ij}}{Z_{ij}^2} & \frac{\omega_0 X_{L,ij}}{Z_{ij}^2} \\
		- \frac{\omega_0 X_{L,ij}}{Z_{ij}^2} & \frac{X_{R,ij}}{Z_{ij}^2}
		\end{bmatrix}, \quad
		H_{i,j}  =  \left[\begin{array}{c;{2pt/2pt}c}
		H_{i,j_{1,1}}  & 0 \\ \hdashline[2pt/2pt]
		0 & 0
		\end{array}\right] % 6 by 6
		\end{align}
	\end{subequations}
	%\hrulefill
	%%%\vspace{-1em}
	% \setcounter{equation}{\value{MYtempeqncnt}}
\end{figure*}

The compositionality results, as well as the verification and synthesis algorithms in Section \ref{sec:seq} can similarly be extended to networks of switched systems. 

\begin{remark}[Extension to nonlinear systems]
In \cite{agarwal2017feedback}, it was shown that the passivity of a nonlinear switched system in a neighborhood around any operating point can be inferred from that of its linear approximation. These results were further extended in \cite{sivaranjani2018mixed} to guarantee the dissipativity of a network of nonlinear switched systems through controllers designed in a centralized manner, using the network formed by linear approximations of the nonlinear switched systems. It is easy to see that similar linear approximation arguments can be used to extend the distributed synthesis procedure proposed in this paper to guarantee dissipativity of a networked system comprised of nonlinear subsystems. Additionally, recently developed notions of equilibrium-independent dissipativity (passivity) \cite{hines2011equilibrium}\cite{simpson2018equilibrium} can be explored in the context of guaranteeing dissipativity around non-trivial operating points. 
\end{remark}

\section{Case Study: Microgrid Network}\label{sec:case}
In this section, we consider the problem of compositional synthesis of local controllers for power networks with large-scale integration of renewables. In such networks, several small distributed generation units (DGUs) and loads are aggregated in clusters known as microgrids. In microgrids, since renewable inputs like wind speed and solar intensity vary continuously, and DGUs either participate or do not participate in the network depending on availability and requirement, switching dynamics are inherent \cite{agarwal2017feedback}. Therefore, it is necessary to synthesize local controllers for DGUs in a compositional manner, such that the stability of the microgrid is maintained when new DGUs connect to the grid, without requiring redesign of existing DGU controllers. In this section, we demonstrate the application of our distributed synthesis framework to enable this `plug-and-play' operation of DGUs in a microgrid.

\begin{figure}[H]
	\centering
	\includegraphics[width=\columnwidth, trim = 0cm 0cm 0cm 0.5cm]{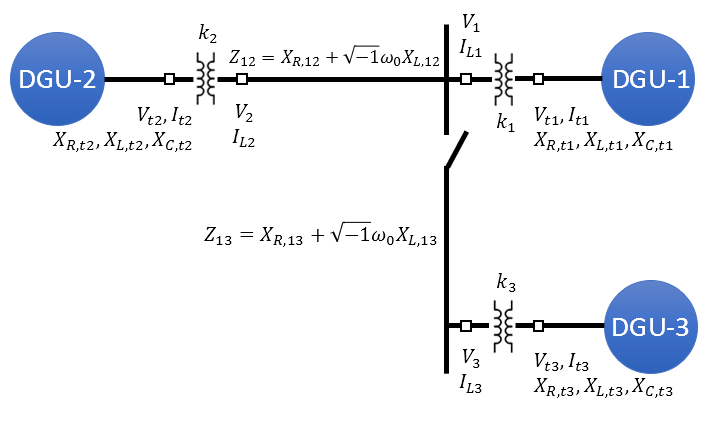}
	\caption{Topology of power network with three DGUs. DGU-3 can connect to and disconnect from DGU-1.}
	\label{fig:microgrid_topology}
	% %%\vspace{-0.1em}
	% \hrulefill
\end{figure}

We consider a microgrid network with three DGUs as shown in Fig. \ref{fig:microgrid_topology}. Each DGU is modeled as a voltage source with internal voltage $V_{ti}$, connected to an RLC-circuit with resistance, inductance and capacitance given by $X_{R,ti}$, $X_{L,ti}$ and $X_{C,ti}$ respectively. The internal voltage at the DGU is stepped up by a transformer with turn ratio $k_i$ to obtain a terminal voltage $V_i$. The line connecting the $i$-th and $j$-th DGUs is assumed to have an impedance $Z_{ij}=X_{R,ij}+\sqrt{-1}\omega_0 X_{L,ij}$, where $X_{R,ij}$ and $X_{L,ij}$ are the resistance and inductance of the line respectively, and $\omega_0$ is the base frequency of the network. 

The dynamics of the microgrid can be modeled as a networked switched system, with system matrices given by \eqref{eq:microgrid_dynamics}. The system parameters are provided in Table \ref{table:parameters} \cite[Appendix C]{riverso2015plug}. As shown in Fig. \ref{fig:microgrid_topology}, DGU-3 can either connect or disconnect to DGU-1 in the network. The dynamics of the $i$-th DGU switches based on the set of DGUs $\mathcal{S}_{i,\sigma_i}$, to which it is connected, as described in \eqref{eq:microgrid_dynamics_switching}. The states (and outputs) of each DGU comprise of the direct and quadrature axis components \cite{anderson2008power} of the terminal voltages (denoted by $V_{i,d}$ and $V_{i,q}$ respectively) and internal currents at the DGU unit (denoted by $I_{ti,d}$ and $I_{ti,q}$ respectively). The control inputs comprise of the direct and quadrature axis internal voltages of the DGU, denoted by $V_{ti,d}$ and $V_{ti,q}$ respectively. The disturbances correspond to the direct and quadrature axis line currents drawn from the DGU, denoted by $I_{Li,d}$ and $I_{Li,q}$ respectively, which vary based on fluctuations in the power sharing between DGUs.

% \usepackage{multirow}
\begin{table}[H]
		\centering
	\caption{Parameters of the microgrid network \cite{riverso2015plug}}
	\label{table:parameters}
	\renewcommand{\arraystretch}{1.5}
	\small{
	\begin{tabular}{lll}
		\hline
		\multirow{3}{*}{DGU-1}           & \begin{tabular}[c]{@{}l@{}}$X_{R,t1}$ ($m\Omega$)\end{tabular} & 1.2    \\
		& \begin{tabular}[c]{@{}l@{}}$X_{L,t1}$ ($\mu H$)\end{tabular}   & 93.7   \\ 
		& \begin{tabular}[c]{@{}l@{}}$X_{C,t1}$ ($\mu F$)\end{tabular}   & 62.86  \\ \hline
		\multirow{3}{*}{DGU-2}           & \begin{tabular}[c]{@{}l@{}}$X_{R,t2}$ ($m\Omega$)\end{tabular} & 1.6    \\   
		& \begin{tabular}[c]{@{}l@{}}$X_{L,t2}$ ($\mu H$)\end{tabular}   & 94.8   \\   
		& \begin{tabular}[c]{@{}l@{}}$X_{C,t2}$ ($\mu F$)\end{tabular}   & 62.86  \\ \hline
		\multirow{3}{*}{DGU-3}           & \begin{tabular}[c]{@{}l@{}}$X_{R,t3}$ ($m\Omega$)\end{tabular} & 1.5    \\   
		& \begin{tabular}[c]{@{}l@{}}$X_{L,t3}$ ($\mu H$)\end{tabular}   & 107.7  \\   
		& \begin{tabular}[c]{@{}l@{}}$X_{C,t3}$ ($\mu F$)\end{tabular}   & 62.86  \\ \hline
		\multirow{4}{*}{Line parameters} & \begin{tabular}[c]{@{}l@{}}$X_{R,12}$ ($\Omega$)\end{tabular}  & 1.1    \\   
		& \begin{tabular}[c]{@{}l@{}}$X_{R,13}$ ($\Omega$)\end{tabular}  & 0.9    \\   
		& \begin{tabular}[c]{@{}l@{}}$X_{L,12}$ ($mH$)\end{tabular}      & 600    \\   
		& \begin{tabular}[c]{@{}l@{}}$X_{L,13}$ ($mH$)\end{tabular}      & 400    \\ \hline
		Transformer turn ratio           & $k_1 = k_2 = k_3$                                              & 0.0435 \\ \hline
		Base frequency                   & \begin{tabular}[c]{@{}l@{}}$\omega_0$ ($Hz$)\end{tabular}      & 60     \\ \hline
	\end{tabular}}
\end{table}

\begin{figure}[!b]
	\centering
	\vspace{0.3em}
	\includegraphics[scale=0.7,trim=0cm 0cm 0.1cm 0.6cm]{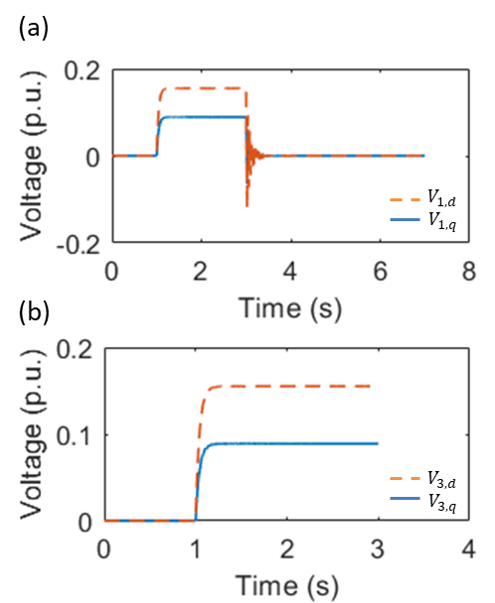}
	\caption{Direct (d) and quadrature (q) axis terminal voltages at (a) DGU-1, and (b) DGU-3.}
	\label{fig:microgrid_plot}
	% %%\vspace{-0.1em}
	% \hrulefill
\end{figure}

Using Theorem \ref{thm:distributed_synthesis_switched}, we design local controllers for all DGUs to guarantee $\mathcal{L}_2$ stability of the network, by choosing $Q_i=-\mathbf{I}$, $S_i=0$ and $R_i=\gamma_i^2 \mathbf{I}$, $i \in \mathbb{N}_3$, where $\gamma_i$ represents the $\mathcal{L}_2$ gain of the closed loop dynamics of $\Sigma_{i}$. The parameters $\gamma_i$ are considered as variables in the synthesis problem $\mathbb{P}_2$, and are found to be
\begin{align*}
\gamma_1&=2.85\\
\gamma_2&=3.21\\
\gamma_3&=3.22.
\end{align*}

We consider a test scenario where DGU-3 connects to the network at $t=1s$ and disconnects at $t=3s$, causing a transient in the system states. The terminal voltage profiles (states) at DGU-1 and DGU-3 during this operation are shown in Fig. \ref{fig:microgrid_plot}, clearly demonstrating that the proposed controllers maintain the stability of the network during plug-and-play operation.

\section{Conclusion}
We presented a distributed and compositional approach to synthesize local controllers for networked systems comprised of dynamically coupled subsystems. The proposed approach can readily be extended to guarantee local dissipativity properties for nonlinear networked systems operating close to equilibrium. Future work will involve extending the distributed synthesis approach to more general classes of nonlinear and hybrid systems.
% Our future work will focus on developing a theory which enables us to add subsystems on both ends of an already existing cascade interconnection in a compositional manner. We will also extend the ideas presented in this paper to more general interconnection topologies of nonlinear systems.

%%%%%%%%%%%%%%%%%%%%%%%%%%%%%%%%%%%%%%%%%%%%%%%%%%%%%%%%
\appendix

\noindent \textbf{Proof of Lemma \ref{lemma:Cholesky_blk}}
\newline
Consider lower triangular matrices ${L}_{i,i}, \ i\in \mathbb{N}_N$ and matrix

\begin{equation}\label{cholesky_decomp_lem}
\begin{aligned}
L&= \begin{bmatrix} L_{i,j} \end{bmatrix}_{i,j\in\mathbb{N}_N}, \\
L_{i,j} &= \begin{cases}
0 &  j>i\\
W_{i,j}L_{j,j}'^{-1} & j \in \{2,\ldots,N\}, \ j<i
\end{cases} \\
L_{i,i}L'_{i,i} \!&=\! \begin{cases} \!W_{i,i} & i =1\\
\!W_{i,i} \!-\! \sum\limits_{j\in\mathbb{N}_{i-1}} L_{i,j}L'_{i,j} & i \in \{2,\ldots,N\} \end{cases}
\end{aligned}
\end{equation}

\noindent with ${W}_{i,j}$, $i,j \in \mathbb{N}_N$ being elements of $W$ as defined in \eqref{eq:block_matrix}. Define ${M}_i = L_{i,i}L'_{i,i}$, $\forall i\in\mathbb{N}_N$. 

A symmetric matrix $M_i$ is positive definite if and only if there exists a lower triangular matrix $L_{i,i}$ with positive diagonal entries such that $M_i = L_{i,i}L'_{i,i}$ \cite[Section 4]{golub2012matrix}. Therefore, if \eqref{seq_Cholesky_lemma} holds, ${L}_{i,i}$ will exist with positive diagonal entries. Invertibility of $L_{i,i}, \, i\in\mathbb{N}_N$ guarantees the existence of $L_{i,j}, \, i,j\in\mathbb{N}_N, \, j<i$. Thus, we can always find a lower triangular matrix $L$ of the form \eqref{cholesky_decomp_lem}, with positive diagonal entries, such that $W = LL'$. This implies the positive definiteness of $W$ \cite[Section 4]{golub2012matrix}, proving the sufficiency of Lemma \ref{lemma:Cholesky_blk}.  Along similar lines, we can also prove the necessity of \eqref{seq_Cholesky} for the positive definiteness of $W$.  \hfill$\blacksquare$

\vspace{1em}

\noindent \textbf{Proof of Theorem \ref{thm:seq_analysis}}
%\newline

\noindent From Proposition \ref{prop:central_analysis}, \eqref{interconnected_linear} is $QSR$-dissipative if
 
  \begin{equation}
    \Gamma = \left [ \begin{array}{c;{2pt/2pt}c}
    -\hat{A}'P-P\hat{A}+C'QC & -PB^{(2)}+C'S \\ \hdashline[2pt/2pt]
    -B^{(2)'}P+S'C & R
    \end{array}\right ]\geq 0,
\end{equation}

\noindent where $\hat{A} = A + B^{(1)}H$. Consider $Q = \mbox{diag}(Q_1,Q_2,\dots, Q_N)$, $S = \mbox{diag}(S_1,S_2,\dots, S_N)$, $R = \mbox{diag}(R_1,R_2,\dots, R_N)$ and $P = \mbox{diag}(P_1,P_2,\dots, P_N)$, where $Q_i\in\mathbb{R}^{m_i \times m_i}$, $S_i\in\mathbb{R}^{m_i \times l_i}$, $R_i\in\mathbb{R}^{l_i \times l_i}$ and $P_i\in\mathbb{R}^{n_i \times n_i}$, $i\in\mathbb{N}_N$. Consider a permutation matrix
  \begin{equation*}
  \delimitershortfall=0pt
  \setlength{\dashlinegap}{2pt}
      \mathbb{E} = \left [\begin{array}{cc:cc:c:cc}
            e_1' & e_{N+1}' & e_2' & e_{N+2}' & \cdots &  e_N' & e_{N+N}'
      \end{array}\right]',
  \end{equation*}
\noindent  where $e_k, \, k\in \mathbb{R}^{+}\backslash\{0\}$, are defined as $ e_i=[e_i^a,0_{n_i,l}]$, and $e_{N+i}=[0_{l_i,n},e_i^b]$, where 
\begin{itemize}
    \item $e_i^a$ is a matrix with dimension $n_i\times n$, that contains all zero elements, but an identity matrix of dimension $n_i$ at columns $(\sum_{j=1}^{i-1} n_j)+1:(\sum_{j=1}^i n_j)$.
    \item $e_i^b$ is a matrix with dimension $l_i\times l$, that contains all zero elements, but an identity matrix of dimension $l_i$ at columns $(\sum_{j=1}^{i-1} l_j)+1:(\sum_{j=1}^i l_j)$.
    \item $0_{r,c}$ is a matrix, with $r$ rows and $c$ columns, whose entries are all $0$.
\end{itemize}
\noindent  Right multiplication of $\Gamma$ with $\mathbb{E}'$ permutes its columns, and a left multiplication with $\mathbb{E}$ permutes its rows.
  
  \begin{align}
      \mathbb{E}\Gamma\mathbb{E}' & = W = \begin{bmatrix} 
   W_{1,1} & W_{1,2} & \dots & W_{1,N} \\
   W_{2,1} & W_{2,2} & \dots & W_{2,N} \\
   \vdots & \vdots & & \vdots \\
   W_{N,1} & W_{N,2} & \dots & W_{N,N} \\
    \end{bmatrix},\\ \nonumber\\
    W_{i,i} & = \left[ \begin{array}{c;{2pt/2pt}c}
    -\hat{A}_i'P_i-P_i\hat{A}_i'
    +C_i'Q_iC_i & -P_iB_i^{(2)}+C_i'S_i \\ \hdashline[2pt/2pt]
    -B_i^{(2)'}P_i+S_i'C_i & R_i
    \end{array} \right], \\ 
    W_{i,j} & = 
    \left[\begin{array}{c;{2pt/2pt}c}
    -\hat{H}_{i,j}-\hat{H}'_{j,i} & 0 \\ \hdashline[2pt/2pt]
    0 & 0
    \end{array}\right],
  \end{align}
  
\noindent for all $i,j\in\mathbb{N}_N, \ j\neq i$, where $\hat{A}_i = A_i + B_i^{(1)}H_i$ and $\hat{H}_{i,j} = P_iB_i^{(1)}H_{i,j}$. 

Note that $\Gamma\geq 0$ if and only if $\mathbb{E}\Gamma\mathbb{E}' \geq 0$. If \eqref{seq_Cholesky} and \eqref{eq:messenger_matrix} hold, then, from Lemma \ref{lemma:Cholesky_blk}, $\mathbb{E}\Gamma\mathbb{E}'>0$ and all conditions in Proposition \ref{prop:central_analysis} are satisfied with $Q = \mbox{diag}(Q_1,Q_2,\dots, Q_N)$, $S = \mbox{diag}(S_1,S_2,\dots, S_N)$, $R = \mbox{diag}(R_1,R_2,\dots, R_N)$ and $P = \mbox{diag}(P_1,P_2,\dots, P_N)$. Therefore, the networked dynamical system $\mathbf{T_N}$ in \eqref{interconnected_linear} is $QSR$-dissipative.  
\hfill $\blacksquare$   

\vspace{1em}

\noindent \textbf{Proof of Theorem \ref{thm:distributed_synthesis}}
%\newline

\noindent The proof follows by applying Theorem \ref{thm:seq_analysis} to the closed loop system,
\begin{align*}
\dot{x}_i(t) & = A_i x_i(t) + B_i^{(1)}v_i(t) + B_i^{(2)} w_i(t) \\ & \ + B_i^{(3)} \sum \limits_{j \in \mathcal{E}_i}K_{i,j}(t)x_j(t), \\
y_i(t) & = C_ix_i(t), \\
v_i(t) & = \sum \limits_{j \in \mathbb{N}_N} H_{i,j} x_{j}(t).
\end{align*}\hfill$\blacksquare$

\vspace{1em}

\noindent \textbf{Proof of Corollary \ref{cor:comp_synthesis}}

\noindent If $\mathbb{P}_3$ is feasible, then Theorem \ref{thm:distributed_synthesis} holds for $\mathbf{T_{N+1}}=\mathbf{T_N}\vert\Sigma_{N+1}$, thus completing the proof.
\hfill $\blacksquare$

\vspace{1em}

\noindent \textbf{Proof of Theorem \ref{thm:distributed_synthesis_switched}}

\noindent Since Definition \ref{def:diss} holds for switched systems \cite{zhao2008dissipativity}, along the lines of proof for Theorem \ref{thm:seq_analysis}, the networked switched system \eqref{interconnected_switched} with $$u_i(t)\!=\!\sum\limits_{j \in \mathcal{E}_i}\!K_{i,j}x_j(t),$$ $i\in \mathbb{N}_N$, is $QSR$-dissipative~if

 \begin{align}
W &= \begin{bmatrix} 
W_{1,1} & W_{1,2} & \dots & W_{1,N} \\
W_{2,1} & W_{2,2} & \dots & W_{2,N} \\
\vdots & \vdots & & \vdots \\
W_{N,1} & W_{N,2} & \dots & W_{N,N} \\
\end{bmatrix} > 0 \end{align}

\noindent holds, where
\begin{align}
W_{i,i} & = \left[ \begin{array}{c;{2pt/2pt}c}
-\hat{A}_i^{\sigma_i'}P_i-P_i\hat{A}_i^{\sigma_i'}
\\+C_i^{\sigma_i'}Q_iC_i^{\sigma_i} & -P_iB_i^{(2)^{\sigma_i}}+C_i^{\sigma_i'}S_i \\ \hdashline[2pt/2pt]
-B_i^{(2)^{\sigma_i'}}P_i+S_i'C_i^{\sigma_i} & R_i
\end{array} \right],\\
W_{i,j} & =
\left[\begin{array}{c;{2pt/2pt}c}
\hat{H}_{j,i}^{\sigma_j'}+\hat{H}_{i,j}^{\sigma_i} & 0 \\ \hdashline[2pt/2pt]
0 & 0
\end{array}\right],
\end{align}

\noindent $\hat{A}_i^{\sigma_i} = A_i^{\sigma_i} + B_i^{(1)^{\sigma_i}}H_i$,  $	\hat{H}_{i,j}^{\sigma_i} = P_i(B_i^{(1)^{\sigma_i}}H_{i,j}+B_i^{(3)^{\sigma_i}}K_{i,j})$ and $i,j~\in~\mathbb{N}_N, \ j\neq i$ and $\sigma_i \in \mathbb{N}_{\eta_i}$. Clearly, if \eqref{close_loop_seq_Cholesky_switched} and \eqref{eq:messenger_matrix_control_switched} hold, then, from Lemma \ref{lemma:Cholesky_blk}, the closed loop networked switched system $\mathbf{T_N^{\sigma}}$ is $QSR$-dissipative.
\hfill $\blacksquare$

\vspace{1em}

\noindent \textbf{Note on Remark \ref{rem:lyapunov}}
%\newline

\noindent Consider a  switched system $\Sigma$
\begin{align*}
\dot{x}(t) & = A^{\sigma(t)}x(t) + B^{(1)^{\sigma(t)}}v(t) + B^{(2)^{\sigma(t)}}w(t)
+ B^{(3)^{\sigma(t)}}u(t)\\
y(t) & = C^{\sigma(t)}x(t)
\end{align*}
which switches arbitrarily between two modes $\sigma(t) \in \{1,2\}$. Suppose $\Sigma$ is  $QSR$-dissipative with multiple energy matrices (or multiple storage functions), that is, if $\sigma(t)=1, \forall t_0\leq t \leq t_1$, $\Sigma$ satisfies the dissipativity inequality
 $$\int_{t_0}^{t_1} \begin{bmatrix}y(\tau) \\ w(\tau)\end{bmatrix}'\begin{bmatrix}Q & S \\ S' & R\end{bmatrix} \begin{bmatrix}y(\tau) \\ w(\tau)\end{bmatrix} d\tau\geq V_1(x(t_1))-V_1(x(t_0)),$$
 where $V_1(x)=x'P_1x$, and if $\sigma(t)=2, \forall t_0\leq t \leq t_1$, $\Sigma$ satisfies
 $$\int_{t_0}^{t_1} \begin{bmatrix}y(\tau) \\ w(\tau)\end{bmatrix}'\begin{bmatrix}Q & S \\ S' & R\end{bmatrix} \begin{bmatrix}y(\tau) \\ w(\tau)\end{bmatrix} d\tau\geq V_2(x(t_1))-V_2(x(t_0)),$$
 where $V_2(x)=x'P_2x$.
 
 If the dynamics of $\Sigma$ switches from mode 1 ($\sigma=1$) to mode 2 ($\sigma=2$) at time $t$, then, $\sigma(t^-)=1$ and $\sigma(t^+)=2$. Then, 
 \begin{equation}\label{eq:multi-storage1}
  \int_{t^-}^{t^+} \begin{bmatrix}y(\tau) \\ w(\tau)\end{bmatrix}'\begin{bmatrix}Q & S \\ S' & R\end{bmatrix} \begin{bmatrix}y(\tau) \\ w(\tau)\end{bmatrix} d\tau\geq V_2(x(t))-V_1(x(t))
 \end{equation}
  must hold. Since $\Sigma$ is dissipative for arbitrary switching, consider a different switching signal where $\Sigma$ switches from  mode 2 ($\sigma=2$) to mode 1 ($\sigma=1$) at time $t$. Then,
   \begin{equation} \label{eq:multi-storage2}
  \int_{t^-}^{t^+} \begin{bmatrix}y(\tau) \\ w(\tau)\end{bmatrix}'\begin{bmatrix}Q & S \\ S' & R\end{bmatrix} \begin{bmatrix}y(\tau) \\ w(\tau)\end{bmatrix} d\tau\geq V_1(x(t))-V_2(x(t)).
  \end{equation}
  must hold. 
  
  Clearly, both \eqref{eq:multi-storage1} and \eqref{eq:multi-storage2} can hold if and only if $V_1(x) = V_2(x)$, that is, the energy matrices $P_1$ and $P_2$ are the same. A similar argument follows for dynamical systems with more than two switching modes. It is therefore not possible to have different energy matrices in different modes for a switched system that is dissipative for arbitrary switching.

\bibliographystyle{IEEEtran}
\bibliography{references}

%\begin{IEEEbiographynophoto}[{\includegraphics[width=1in,height=1.25in,clip,keepaspectratio]{mshell}}]{Michael Shell}
\vspace{10em}
\begin{IEEEbiography}[{\includegraphics[width=1in,height=1.25in,clip,keepaspectratio]{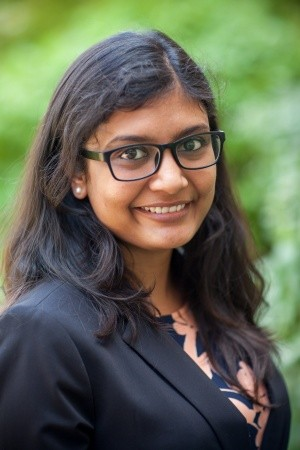}}]{Etika Agarwal}
obtained her PhD in Electrical Engineering from the University of Notre Dame in 2019. She received her Masters in Electrical Engineering from the University of Notre Dame in 2016, and her B. Tech in Avionics from the Indian Institute of Space Science and Technology in 2012. Before joining the graduate school, she worked with the Indian Space Research Organization from 2012-2014. Her research interests are in computationally efficient and scalable control of large-scale cyber-physical systems. 
\end{IEEEbiography}
%\vspace{-5em}
\begin{IEEEbiography}[{\includegraphics[width=1in,height=1.25in,clip,keepaspectratio]{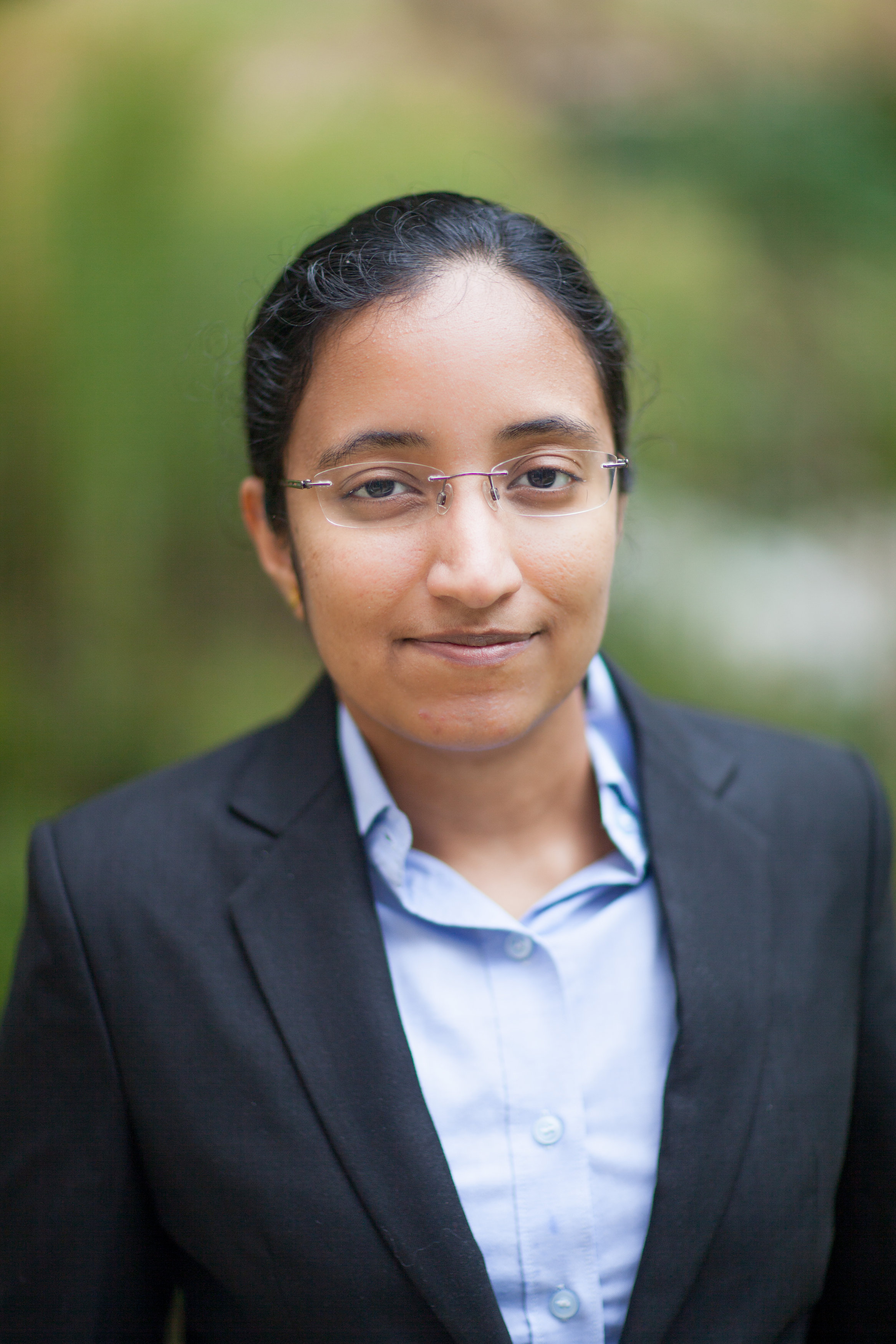}}]{S Sivaranjani}
obtained her PhD in Electrical Engineering from the University of Notre Dame in 2019. She obtained her undergraduate and Master’s degrees in Electrical Engineering from the PES Institute of Technology and the Indian Institute of Science, in 2011 and 2013, respectively. Her research interests are in the area of distributed control for large-scale infrastructure networks, with emphasis on transportation networks and power grids. She is a recipient of the Schlumberger Foundation Faculty for the Future fellowship (2015-2018), the Zonta International Amelia Earhart fellowship (2015-2016) and the Notre Dame (NSF) Ethical Leaders in STEM fellowship (2016-2017).
\end{IEEEbiography}

%\newpage

\begin{IEEEbiography}[{\includegraphics[width=1in,height=1.25in,clip,keepaspectratio]{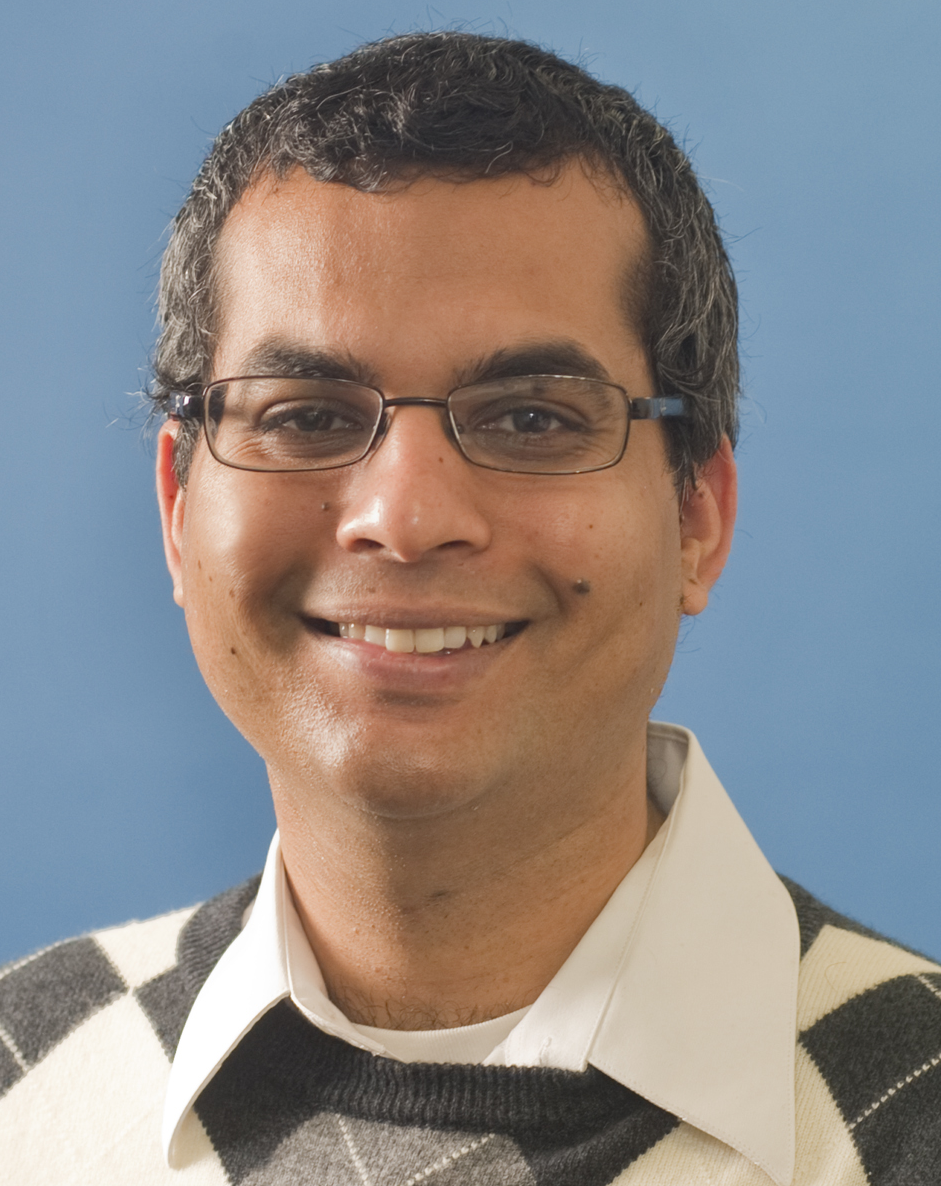}}]{Vijay Gupta}
is a Professor in the Department of Electrical Engineering
	at the University of Notre Dame, having joined the faculty in January
	2008. He received his B. Tech degree at Indian Institute of
	Technology, Delhi, and his M.S. and Ph.D. at California Institute of
	Technology, all in Electrical Engineering. Prior to joining Notre
	Dame, he also served as a research associate in the Institute for
	Systems Research at the University of Maryland, College Park. He
	received the 2018 Antonio Ruberti Award from IEEE Control Systems
	Society, the 2013 Donald P. Eckman Award from the American Automatic
	Control Council and a 2009 National Science Foundation (NSF) CAREER
	Award. His research and teaching interests are broadly at the
	interface of communication, control, distributed computation, and
	human decision making.
\end{IEEEbiography}

\begin{IEEEbiography}[{\includegraphics[width=1in,height=1.25in,clip,keepaspectratio]{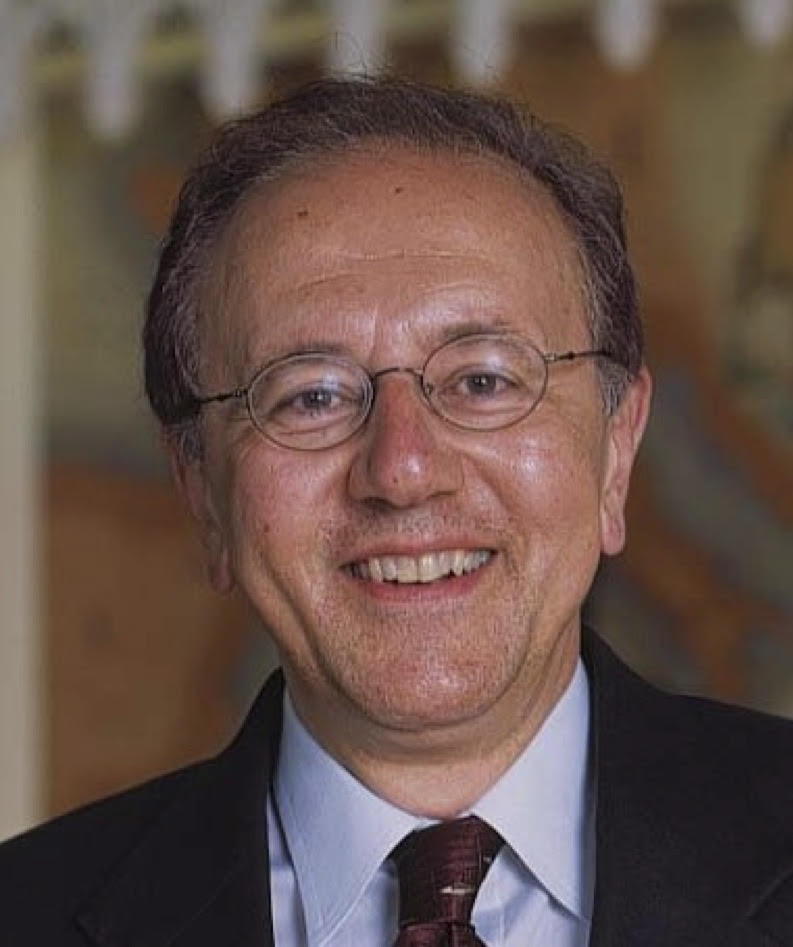}}]{Panos Antsaklis}
is the H.C. \& E.A. Brosey Professor of Electrical Engineering at the University of Notre Dame. He is graduate of the National Technical University of Athens, Greece, and holds MS and PhD degrees from Brown University. His research addresses problems of control and automation and examines ways to design control systems that will exhibit high degree of autonomy. His current research focuses on Cyber-Physical Systems and the interdisciplinary research area of control, computing and communication networks, and on hybrid and discrete event dynamical systems. He is IEEE, IFAC and AAAS Fellow, President of the Mediterranean Control Association, the 2006 recipient of the Engineering Alumni Medal of Brown University and holds an Honorary Doctorate from the University of Lorraine in France. He served as the President of the IEEE Control Systems Society in 1997 and was the Editor-in-Chief of the IEEE Transactions on Automatic Control for 8 years, 2010-2017.
\end{IEEEbiography}

\balance

\end{document}